\documentclass{article}

\usepackage{abstract}
\usepackage{amsmath}
\usepackage{graphicx}
\usepackage{float}
\usepackage{amssymb}

\begin{document}

\title{Q penalties due to pump phase modulation and pump RIN in fiber optic
parametric amplifiers with non-uniform dispersion}

\author{
J.M. Chavez Boggio,
A. Guimar\~aes,
F.A. Callegari,
J.D. Marconi,
H.L. Fragnito
}

\date{}

\maketitle

\noindent
Optics and Photonics Research Center, Instituto de F\'isica
``Gleb Wataghin'', State University of Campinas -- Unicamp,
13083-970, CP 6165, Campinas, SP, Brazil.

\vspace{1em}

\begin{abstract}

We present an experimental and numerical investigation of the quality
factor degradation ($\Delta Q$) due to pump phase modulation and pump
relative intensity noise (RIN) in single- and double-pumped fiber
optical parametric amplifiers (FOPAs). These penalties are
investigated in FOPAs constructed with fibers having constant and
randomly varying zero-dispersion wavelength ($\lambda_0$) along the
fiber.

We show that pump phase modulation in uniform fibers (constant
$\lambda_0$) produces large $Q$ penalties, while in fibers with
variations of $\lambda_0$, these penalties can be strongly reduced in
both single- and double-pumped FOPAs. We also show that $Q$ penalties
due to pump phase modulation tend to disappear in FOPAs made with very
short fibers ($L<0.1~\mathrm{km}$) and very low dispersion slope $
S_0 < 0.002~
\mathrm{ps}\,
\mathrm{nm}^{-2}\,
\mathrm{km}^{-1}.
$ The $Q$ penalties due to pump RIN, which for typical FOPA parameters
are much smaller than those due to pump phase modulation, are also
reduced in non-uniform fibers (varying $\lambda_0$), but do not depend
on fiber length nor on fiber dispersion slope.

\end{abstract}

\bigskip

\noindent

\bigskip

\noindent
DOI:
\texttt{10.1016/j.optcom.2005.01.056}

\bigskip

\noindent
*Corresponding author.*

Tel.: +55 19 788 5451

Fax: +55 19 788 5427

E-mail:
\texttt{jmchavez@ifi.unicamp.br}
(J.M. Chavez Boggio)

\section{Introduction}

Fiber optic parametric amplifiers (FOPAs) have attracted great interest
because of their multiple applications in signal amplification,
wavelength conversion, and optical signal processing
[1,2,3,4,5,6,7,8,9,10,11,12,13,14,15,16,17,18,19,20].
A drawback of these devices is the sensitivity to time-varying pump
frequency or pump power. In most experiments on FOPAs, the pump linewidth is intentionally
broadened through phase modulation to suppress stimulated Brillouin
scattering (SBS) in the fiber. However, as observed in
[21,22,23,24], the variation of the instantaneous pump frequency
due to phase modulation causes instantaneous gain variations that impair
the performance of FOPAs.

On the other hand, the relative intensity noise (RIN) from pump lasers
can be transferred to the signal through the variations of the
parametric gain [25,26,27]. Furthermore, in present experiments
with FOPAs, erbium-doped fiber amplifiers (EDFAs) are used to boost the
pump lasers. Amplified spontaneous emission (ASE) noise from EDFAs
beats with the pump laser, providing an additional source of RIN. The
amount of ASE-induced RIN depends on the noise figure of the booster
EDFA and on the input RIN of the pump laser, and must be considered in
the design of FOPA experiments [25].

In this paper we study, experimentally and numerically, the degradation
of the signal quality factor ($Q$) produced by gain fluctuations caused
by pump frequency or power variations in single-pumped (1P-FOPA) and
double-pumped (2P-FOPA) devices. This investigation considers FOPAs
constructed with both uniform and non-uniform (varying
zero-dispersion wavelength) fibers, and compares their performances.

In Section~2 we describe the formalism used to analyze the $Q$
penalties due to pump variations. In Sections~3 and~4 we present our
numerical results for the 1P-FOPA and the 2P-FOPA, respectively. In
Sections~5 and~6 we present our experimental work. Finally, in
Section~7 we draw our conclusions.

\section{Analysis of Q factor degradation}

We consider a digital non-return-to-zero (NRZ) system in which the
levels one and zero are detected with average electrical currents
$\langle I_1\rangle$ and $\langle I_0\rangle$, and standard deviations
$\sigma_1$ and $\sigma_0$, respectively. The performance degradation of
the system can be estimated from the quality factor

\begin{equation}
Q=
\frac{\langle I_1\rangle-\langle I_0\rangle}
{\sigma_1+\sigma_0}.
\tag{1}
\end{equation}

We assume that $\langle I_0\rangle=0,$ and that the noise in level zero is due only to
spontaneous--spontaneous beat noise and is therefore negligible
[28]. (The small differences introduced by
$\sigma_0\neq0$ are considered in Appendix~A.)

With these assumptions, the quality factor in the case of an ideal
FOPA without gain fluctuations is

\begin{equation}
Q_i=
\frac{\langle I_1\rangle}
{\sigma_1}.
\tag{2}
\end{equation}

In actual FOPAs, gain fluctuations induce variations in the detected
level one, which is assumed to be a Gaussian random variable. Because
of the sub-femtosecond response time of the nonlinear refractive index
in silica fibers, only level one is affected. Spontaneous parametric
fluorescence (i.e., spontaneous annihilation of pump photons creating
signal--idler photon pairs), which introduces noise in level zero, is
negligible in practice. Therefore the quality factor is degraded from
$Q_i$ to

\begin{equation}
Q_n=
\frac{\langle I_1\rangle}
{\sqrt{\sigma_1^2+\sigma_n^2}},
\tag{3}
\end{equation}

where $\sigma_n$ is the standard deviation of the fluctuations in level
one due to signal gain variations in the FOPA. It is proportional to
the standard deviation of the output signal power,
$\delta P_s$.

The $Q$ penalty is defined as $\Delta Q=\frac{Q_i}{Q_n},$ which yields

\begin{equation}
\Delta Q=
\sqrt{
1+
\frac{\sigma_n^2}
{\langle I_1\rangle^2}
Q_i^2
}.
\tag{4}
\end{equation}

Since $I_1$ is proportional to the output signal power $P_s$, and $\sigma_n^2$ is proportional to the variance of the output optical
power $\langle \delta P_s^2\rangle$, with $\langle \delta P_s^2\rangle=\sigma_G^2,$ the relative intensity signal noise due to signal gain variations is approximated by [19]

\begin{equation}
\frac{\sigma_n^2}{\langle I_1\rangle^2}
=
\frac{\langle \delta P_s^2\rangle}
{\langle P_s\rangle^2}
=
\left(
\frac{\sigma_G}
{\langle G\rangle}
\right)^2.
\tag{5}
\end{equation}

where $\sigma_G$ and $\langle G\rangle$ are the standard deviation and
mean of the parametric gain, respectively.

Replacing Eq.~(5) into Eq.~(4), the $Q$ penalty becomes

\begin{equation}
\Delta Q(\mathrm{dB})
=
10\log_{10}
\left[
\sqrt{
1+
Q_i^2
\left(
\frac{\sigma_G}
{\langle G\rangle}
\right)^2
}
\right].
\tag{6}
\end{equation}

The $Q$ penalty therefore depends on the baseline performance $Q_i$.
Throughout this paper we assume $Q_i=11$, which is a typical value and
corresponds to our experiments.

To calculate $\sigma_G/\langle G\rangle$ in FOPAs we follow two different approaches. The simpler one consists of using the well-known analytical expressions
for $G$ to calculate the partial derivative of the gain with respect to
the pump power or pump frequency. This approach, detailed in Appendix
B, leads to analytical expressions that provide physical insight, but
does not take into account longitudinal variations of the
zero-dispersion wavelength, which are important in practical FOPAs.

In the second approach, adopted throughout this paper, we solve
numerically the signal propagation equation for a set of pump
frequencies or pump powers that simulate temporal variations of these
quantities. The average and standard deviation of the parametric gain
are then calculated for each signal wavelength. This procedure is
justified by the ultrafast response of the Kerr nonlinearity, so that,
if the pump fluctuations follow ergodic random processes, averages over
time coincide with ensemble averages over a set of fixed pump
parameters.

In the frequency domain, the coupled propagation equations for the
signal at frequency $\omega_s$ and the idler at frequency $\omega_i$
are [8,29].

\begin{align}
\frac{\partial A_s}{\partial z}
&=
a(z)A_i^{*},
\tag{7a}\\
\frac{\partial A_i}{\partial z}
&=
a(z)A_s^{*}.
\tag{7b}
\end{align}

Here $z$ is the position along the fiber (total length $L$),
$A_s$ and $A_i$ are proportional to the signal and idler field
amplitudes, the superscript $*$ denotes complex conjugation, and the
coefficient $a$ depends on whether the amplifier is single- or
double-pumped.

Equations (7) can be decoupled, yielding

\begin{equation}
\frac{\partial^2A_s}{\partial z^2}
+
b
\frac{\partial A_s}{\partial z}
-
|a|^2A_s
=
0,
\tag{8}
\end{equation}

where $b=-\frac{d\ln(a)}{dz}.$ For the single-pump FOPA, the coefficients are

\begin{align}
a_1(z)
&=
-i\gamma P_1
\exp\!\left[
i\left(
2\phi+\Delta\beta z
\right)
\right],
\tag{9a}
\\
b_1
&=
\alpha
-
i\left(
\Delta\beta
+
2\gamma P_1e^{-\alpha z}
\right).
\tag{9b}
\end{align}

Here $\alpha$ is the fiber attenuation, $\gamma$ is the nonlinear coefficient, $P_1$ is the pump power, and $
\Delta\beta
=
\beta(\omega_s)
+
\beta(\omega_i)
-
2\beta(\omega_1),
$ while $
\phi(z)
=
\phi_1
+
\frac{\gamma P_1
\left(
1-e^{-\alpha z}
\right)}
{\alpha},
$ where $\omega_1$ is the pump frequency and $\phi_1$ is the initial pump
phase.

For the double-pump FOPA, the coefficients become

\begin{align}
a_2
&=
-2i\gamma
\sqrt{P_1P_2}\,
\exp\!\left[
i\Delta\beta z
+i\phi
-\alpha z
\right],
\tag{10a}
\\
b_2
&=
\alpha
-
i\Delta\beta
+
\gamma(P_1+P_2)e^{-\alpha z}.
\tag{10b}
\end{align}

Here $P_1$ and $P_2$ are the input pump powers, $
\Delta\beta
=
\beta(\omega_s)
+
\beta(\omega_i)
-
\beta(\omega_1)
-
\beta(\omega_2),
$ and $
\phi
=
\phi_1+\phi_2
+
\frac{\gamma(P_1+P_2)
\left(
1-e^{-\alpha z}
\right)}
{\alpha},
$ where $\phi_j$ is the initial phase of the $j$th pump laser. The output signal power is $
P_s(L)=|A_s(L)|^2,
$ and the signal gain is $
G=\frac{P_s(L)}{P_s(0)},
$ which is obtained by solving Eq.~(8) with the matrix method described
in Ref.[8].

The probability distribution describing the pump-frequency variation
depends on the phase-modulation technique. In the most commonly used
approach [1,5], the phase modulator is driven by three or four
electrically combined sinusoidal RF signals whose frequencies are chosen
to suppress stimulated Brillouin scattering. Experimentally, when SBS
is efficiently suppressed, the optical spectrum of the pump is nearly
flat-topped. Therefore, throughout this paper the pump-frequency
variations are modeled by a uniform distribution, as discussed in
Sections~3.1 and~4.1. (See Appendix~C for a discussion of other phase
modulation methods.)

We simulated the stochastic pump-power variations through a large
number ($N=800$) of pump-power values randomly generated from a
Gaussian probability distribution with standard deviation $\sigma_P$.
This process generates a set of $N$ gain spectra. For each signal
frequency we calculate the average gain,
$\langle G\rangle$, and its standard deviation,
$\sigma_G$, which are then used in Eq.~(6) to obtain the
$Q$ penalty.

\section{Numerical results: 1P-FOPA}

The following parameters were employed to numerically solve Eq.~(8): $
P_1 = 330~\mathrm{mW}, 
$ $L = 6.8~\mathrm{km},
$ $\alpha = 0.25~\mathrm{dB/km},
$ $\gamma = 2~(\mathrm{W\,km})^{-1},
$ $\lambda_0 = 1567~\mathrm{nm},
$ where $\lambda_0=\frac{2\pi c}{\omega_0},
$ the third-order dispersion is $
\beta_3(\omega_0)
=
0.11~
\mathrm{ps}^3/\mathrm{km},
$ and the fourth-order dispersion is $
\beta_4(\omega_0)
=
-0.0004~
\mathrm{ps}^4/\mathrm{km}.
$

Throughout this paper the third- and fourth-order dispersion
coefficients are calculated from
$\beta_3(\omega_0)
=
\frac{4\pi^2c^2S_0}
{\lambda_0^4},$ 
and $
\beta_4(\omega_0)
=
-
\frac{16\pi^2c^2S_0}
{\lambda_0^5},
$ where $
S_0
=
0.074~
\mathrm{ps}
\,
\mathrm{nm}^{-2}
\,
\mathrm{km}^{-1}
$ is the fiber dispersion slope. For the numerical solution of Eq.~(8), the fiber is divided into
3000 steps (approximately $2.2~\mathrm{m}$ long), where
$|a|^2$ and $b$ are assumed constant within each step
(but vary from one step to the next because of pump-power loss
and variations of $\lambda_0$ along the fiber).

\subsection{Pump instantaneous frequency variation}

The pump frequency $\nu_1$ is assumed to follow a uniform
distribution over a bandwidth of $5~\mathrm{GHz}$ (corresponding to approximately $
0.04~\mathrm{nm}),$ which is a typical value used to suppress stimulated Brillouin
backscattering in experiments with FOPAs. The frequency variation is simulated using $N=400$ pump wavelengths uniformly spaced with a separation of $0.0001~\mathrm{nm}.
$

\subsubsection{Case 1: Uniform fiber (constant $\lambda_0$)}

Figure~1(a) shows the calculated $Q$ penalty,
$\Delta Q$, together with the average gain spectrum,
$\langle G\rangle$, as functions of the signal wavelength,
when the average pump wavelength satisfies $
\lambda_1
=
\lambda_0
+
0.1~\mathrm{nm}.
$ The $Q$ penalty approaches zero only for signal wavelengths located
near the pump wavelength or around the gain maxima.

Very large penalties
(approximately $
\Delta Q \approx 10~\mathrm{dB}
$ for the parameters used here)
occur at the outer slopes of the gain spectrum,
leading to a complete degradation of the 1P-FOPA performance.

A second maximum, $\Delta Q \approx 3~\mathrm{dB},
$ appears near the inflection points of the gain spectrum located
between the pump wavelength and the gain peaks,
as previously observed in Ref.~24. Unfortunately, this second degradation peak occurs in the spectral
region where the gain is highest.

Throughout the remainder of the paper the largest value of
$\Delta Q$ will be referred to as the
\emph{first peak},
while the smaller maximum will be referred to as the
\emph{second peak}
(see Fig.~1(a)).

For comparison, Fig.~1(a) also includes the analytical prediction of
the gain (segmented curve) and of $\Delta Q$ (dotted curve), obtained
using the average pump power along the fiber, $
\bar P_1
=
P_1
\frac{1-e^{-\alpha L}}
{\alpha L}.
$ The analytical and numerical gain spectra agree very well. For $\Delta Q$, however, the analytical model overestimates the
penalty by approximately $2~\mathrm{dB}$ at the first peak and $
0.2~\mathrm{dB}
$ at the second peak for the parameters considered in Fig.~1.

\bigskip

The results indicate that when the pump frequency $\nu_1$ varies, the
performance of the FOPA degrades most strongly in those spectral
regions where the gain spectrum changes most rapidly with the signal
wavelength $\lambda_s$, i.e., around (although not exactly at) the
inflection points of the gain spectrum. Conversely, the FOPA exhibits
its best performance for signal wavelengths located near the extrema
(maxima or minima) of the gain spectrum. It should be noted, however, that the maxima of $\Delta Q$ do not
coincide exactly with the inflection points of the gain spectrum,
because the derivatives of the gain with respect to the pump wavelength
and to the signal wavelength are different (see Appendix~B).

We now investigate the influence of tuning the average pump wavelength
closer to, or farther from, the zero-dispersion wavelength. It is well known that the bandwidth of a single-pump FOPA increases as
the pump wavelength approaches $\lambda_0$, reaching its maximum when $
\lambda_1 \approx \lambda_0.
$ However, as shown in Fig.~1(b), the $Q$ penalty increases as the pump
approaches the zero-dispersion wavelength. This behavior can be understood from the fact that small variations in
$\lambda_1$ produce much larger changes in the shape of the gain
spectrum when $\lambda_1 \approx \lambda_0 .
$ Figure~1(b) shows that, in order to keep the second-peak penalty below
a prescribed upper bound, the separation
$\lambda_1-\lambda_0$ must remain larger than a critical value.

For the parameters employed in our simulations, maintaining $
\Delta Q < 0.5~\mathrm{dB}
$ requires $
\lambda_1-\lambda_0 > 0.35~\mathrm{nm}.
$ Therefore, in general there is a trade-off between gain bandwidth and
system penalty in single-pump FOPAs employing a phase-modulated pump. Increasing the pump power $P_1$ increases the parametric gain, but the
gain fluctuations increase even faster. Consequently, the $Q$ penalty
increases monotonically. The inset of Fig.~1(b) shows the values of the first and second
$\Delta Q$ peaks as a function of the peak gain (expressed in dB),
which, for large gains, is approximately proportional to the pump
power $P_1$. At the second peak the penalty reaches approximately $
\Delta Q \approx 5~\mathrm{dB}
$ for peak gains close to $
50~\mathrm{dB}.
$ 

We next investigate the influence of the fiber length $L$ on the
penalty produced by pump phase modulation. The fiber length $L$ and the nonlinear coefficient $\gamma$ were varied
simultaneously in order to keep the peak gain constant. Figure~2 shows that for fiber lengths larger than approximately $L \gtrsim 4~\mathrm{km},$ the penalty depends only weakly on the fiber length; increasing $L$
produces only a slight increase in $\Delta Q$. On the other hand, decreasing the fiber length reduces the penalty. For very short fibers, $
L < 0.1~\mathrm{km},
$ the reduction becomes pronounced, and the second peak of $\Delta Q$
approaches zero. Since the penalties caused by pump phase modulation depend on the
proximity between the pump wavelength and the zero-dispersion
wavelength, it is natural to expect a dependence on the fiber
dispersion slope, as previously pointed out in Ref. [23].

The dispersion slope was reduced from $
S_0
=
0.074~
\mathrm{ps}\,
\mathrm{nm}^{-2}\,
\mathrm{km}^{-1}
$ to $
S_0
=
0.002~
\mathrm{ps}\,
\mathrm{nm}^{-2}\,
\mathrm{km}^{-1},
$ and the corresponding values of $\Delta Q$ were calculated.

The inset of Fig.~2 shows that the first and second penalty peaks both
decrease as $S_0$ decreases. For $
S_0
\approx
0.002~
\mathrm{ps}\,
\mathrm{nm}^{-2}\,
\mathrm{km}^{-1},
$ the second peak becomes negligible, $
\Delta Q < 0.2~\mathrm{dB}.
$

\bigskip

The results shown in Fig.~2 indicate that the $Q$ penalties produced by
pump phase modulation can be greatly reduced by simultaneously using $
L < 0.1~\mathrm{km}
$ and $
S_0 < 0.01~
\mathrm{ps}\,
\mathrm{nm}^{-2}\,
\mathrm{km}^{-1}.
$ The analytical treatment presented in Appendix~B leads to the same
conclusion. Furthermore, Appendix~B also shows that, for very short
fibers ($L<0.1~\mathrm{km}$), the $Q$ penalties become much less
sensitive to the separation between the pump wavelength and the
zero-dispersion wavelength than in long fibers.

\subsubsection{Case 2: Non-uniform fiber (varying $\lambda_0$)}

We now investigate the influence of longitudinal fluctuations of the
zero-dispersion wavelength on the performance of the FOPA. In practical fibers, the zero-dispersion wavelength is not constant,
but varies randomly along the propagation direction
[8,9,10,33,34,35,36,37,38,39,40]. The fiber considered here exhibits random fluctuations of
$\lambda_0(z)$ over both short length scales (of the order of meters)
and long length scales (hundreds of meters). The statistical model used
to generate these fluctuations is described in Appendix~D. Figure~3(a) shows the second-peak $Q$ penalty as a function of the
standard deviation of the zero-dispersion wavelength, $
\sigma_{\lambda_0},
$ for $
\lambda_1-\langle\lambda_0\rangle
=
0.1~\mathrm{nm}
$ and $
\lambda_1-\langle\lambda_0\rangle
=
0.2~\mathrm{nm}.
$

To obtain reliable statistics, each point corresponds to the average of
15 different fiber realizations, each having different longitudinal
variations of $\lambda_0(z)$ but identical values of $\langle\lambda_0\rangle$ and $\sigma_{\lambda_0}$. The error bars indicate the spread of the results obtained from the 15
simulations. The average $Q$ penalty initially increases for small values of $\sigma_{\lambda_0}$, reaches a maximum, and then decreases. Remarkably, when $
\sigma_{\lambda_0}
>
0.07~\mathrm{nm},
$ for the case $
\lambda_1-\langle\lambda_0\rangle
=
0.1~\mathrm{nm},
$ the penalty becomes smaller than in the perfectly uniform fiber
($\sigma_{\lambda_0}=0$). This reduction of the $Q$ penalty for sufficiently large
$\sigma_{\lambda_0}$ can be explained by the fact that large
fluctuations of the zero-dispersion wavelength create fiber segments
for which $
\left|
\lambda_1-\lambda_0
\right|
$ is relatively large. These fiber sections still contribute to the overall gain but are much
less sensitive to variations of the pump wavelength. Furthermore, longitudinal variations of $\lambda_0$ smooth the gain
spectrum, which also contributes to reducing the $Q$ penalty.

We next investigate another type of longitudinal variation of the
zero-dispersion wavelength. The fiber still contains the same short-scale random fluctuations
$\delta\lambda_0(z)$ with standard deviation approximately $
0.02~\mathrm{nm},
$ but the long-scale variation is no longer random. Instead, the fiber is divided into three equal sections, each of length $
\frac{L}{3},
$ whose zero-dispersion wavelengths satisfy $
\lambda_{01}
=
1567~\mathrm{nm}
-
\frac{\Delta\lambda_0}{2},
$ $
\lambda_{02}
=
1567~\mathrm{nm},
$ $
\lambda_{03}
=
1567~\mathrm{nm}
+
\frac{\Delta\lambda_0}{2}.
$ For $
\Delta\lambda_0
\gg
0.02~\mathrm{nm},
$ the corresponding standard deviation is $
\sigma_{\lambda_0}
=
\frac{\Delta\lambda_0}{\sqrt{6}}.
$ Figure~3(b) shows the second-peak $Q$ penalty as a function of
$\sigma_{\lambda_0}$. Small values of $\Delta\lambda_0$ degrade the system performance. However, for sufficiently large values of
$\Delta\lambda_0$, the performance improves and becomes even better
than that of the perfectly uniform fiber ($\Delta\lambda_0=0$ and $\delta\lambda_0(z)=0$). It is also observed that increasing $
\lambda_1-\langle\lambda_0\rangle
$ reduces the $Q$ penalty and decreases its sensitivity to $\Delta\lambda_0$.

\bigskip

The improvement in the $Q$ penalty obtained for large values of $\sigma_{\lambda_0}$ does not depend on the particular longitudinal
distribution of $\lambda_0(z)$. Furthermore, we verified that this improvement is a general property of
fiber optical parametric amplifiers and is also observed for different
fiber lengths, nonlinear coefficients, attenuation coefficients, and
pump powers. This result has important implications when different values of $\lambda_0$ are deliberately arranged to produce dispersion-tailored
fibers with broadband and nearly flat gain spectra, as proposed in Ref.[7]. As an example, we generated the gain spectrum shown in Fig.~4 using the
following fiber parameters: $
\gamma
=
20~(\mathrm{W\,km})^{-1},
$ $
P_1
=
0.65~\mathrm{W},
$ $
L
=
0.42~\mathrm{km},
$ $
\beta_3(\omega_0)
=
0.033~
\mathrm{ps}^3/\mathrm{km},
$ $
\beta_4(\omega_0)
=
0.0003~
\mathrm{ps}^4/\mathrm{km},
$ and $
\alpha
=
0.058~\mathrm{km}^{-1}.
$ The fiber consists of six segments (without short-range fluctuations)
whose zero-dispersion wavelengths and fractional lengths are listed in
Table~1. The resulting gain spectrum exhibits a gain ripple smaller than $
1.5~\mathrm{dB}
$ over a bandwidth of approximately $
45~\mathrm{nm},$ while the corresponding $Q$ penalty remains below $0.6~\mathrm{dB}
$ throughout this region. These examples demonstrate that FOPAs combining short fiber lengths, low dispersion slope, large longitudinal variations of $\lambda_0$, and a suitable arrangement of the dispersion profile, can simultaneously provide low system penalties, broadband operation, and good gain flatness.

\subsection{Stochastic pump power variation}

Typical distributed-feedback (DFB) lasers suitable for use as pump
sources exhibit relative intensity noise (RIN) of approximately $
-170~\mathrm{dB/Hz}.$ When the pump is amplified by an erbium-doped fiber amplifier (EDFA),
its RIN typically increases by about $10~\mathrm{dB},
$ leading to practical values in the range $-160
\ \text{to}\
-150~\mathrm{dB/Hz}.$ A RIN of $
-160~\mathrm{dB/Hz},
$ integrated over a detection bandwidth of $
7~\mathrm{GHz},$ corresponds to stochastic pump-power fluctuations $
\frac{\sigma_{P_1}}
{\langle P_1\rangle}
=
8.35\times10^{-4},
$ whereas $
-150~\mathrm{dB/Hz}$ corresponds to $\frac{\sigma_{P_1}}
{\langle P_1\rangle}
=
2.65\times10^{-3}.
$ In this section we investigate the $Q$ penalties produced by pump RIN
for fibers with and without longitudinal variations of the
zero-dispersion wavelength.

\subsubsection{Case 1: Uniform fiber (constant $\lambda_0$)}

Figure~5(a) presents the calculated spectra of $\Delta Q$ and $\langle G\rangle$
for pump RIN values of $
-160~\mathrm{dB/Hz}
$ and $
-150~\mathrm{dB/Hz},
$ assuming $
\lambda_1-\lambda_0
=
0.1~\mathrm{nm}.
$ As expected, the larger RIN produces larger
$Q$ penalties. As the signal wavelength varies, the penalty approaches zero near the
pump wavelength, increases rapidly with increasing $
|\lambda_1-\lambda_s|,
$ and reaches its maximum near (although not exactly at) the inflection
points located on the outer slopes of the gain spectrum (see
Appendix~B). For comparison, the analytical prediction obtained using the average
pump power $
\bar P_1
=
P_1
\frac{1-e^{-\alpha L}}
{\alpha L}
$ is also shown. Only a small quantitative disagreement is observed between the
analytical and numerical calculations.

A comparison with the phase-modulation case shows that the penalties
produced by pump RIN are considerably smaller. For identical fiber and pump parameters, $
\Delta Q_{\max}
<
0.65~\mathrm{dB}
$ for pump RIN, whereas $
\Delta Q_{\max}
\approx
10.5~\mathrm{dB}
$ for pump phase modulation.

\bigskip

\noindent
\textbf{Table 1.}

Zero-dispersion wavelength distribution used for the flat-gain
single-pump FOPA.

The table lists the value of $\lambda_0$ for each fiber segment,
its corresponding length, together with the overall mean value
$\langle\lambda_0\rangle$ and standard deviation
$\sigma_{\lambda_0}$. For larger values of pump RIN, a representative value of the
$Q$ penalty is obtained by evaluating $\Delta Q$ at the signal
wavelength corresponding to the gain peak. Figure~5(b) shows this quantity as a function of the pump RIN. It can be observed that, in order to maintain $\Delta Q < 0.2~\mathrm{dB},$ which is a tolerable value for practical communication systems, the pump RIN must satisfy $ \mathrm{RIN}
<
-145~\mathrm{dB/Hz}.$ From our numerical simulations we also found that the maximum
$Q$ penalty produced by pump RIN is practically independent of the
pump detuning $
|\lambda_1-\lambda_0|.
$ On the other hand, increasing the pump power (and consequently the
parametric gain), while keeping the pump RIN constant, increases the
resulting $Q$ penalty. A similar behavior has previously been reported for the noise figure
of single-pump FOPAs [27].

\subsubsection{Case 2: Non-uniform fiber (varying $\lambda_0$)}

To investigate the influence of longitudinal fluctuations of the
zero-dispersion wavelength, we employed the same statistical procedure
used previously for the phase-modulation analysis (Appendix~D). The simulations were performed using $
\lambda_1-\langle\lambda_0\rangle
=
0.1~\mathrm{nm},
$ a pump RIN of $
-150~\mathrm{dB/Hz},
$ while varying the standard deviation $
\sigma_{\lambda_0}.
$

Figure~6(a) presents the average value of the $Q$ penalty,
evaluated at the gain peak wavelength, $
\langle\Delta Q\rangle,$ as a function of
$\sigma_{\lambda_0}$. For comparison, the same figure also shows the evolution of the
average peak gain. The calculations indicate that both the gain and the corresponding
$Q$ penalty decrease as $\sigma_{\lambda_0}$ increases,
eventually reaching an approximately constant value for $
\sigma_{\lambda_0}
\simeq
0.05~\mathrm{nm}.
$ The reduction of the $Q$ penalty can be understood by examining the
typical spectra shown in Fig.~6(b), obtained for $\sigma_{\lambda_0}
=
0.021~\mathrm{nm}.
$ Compared with the spectrum of the uniform fiber, the gain profile
becomes smoother, especially at its outer slopes. Since the gain varies more slowly with the signal wavelength,
the sensitivity of the amplifier to pump-power fluctuations is reduced,
which directly lowers the corresponding $Q$ penalty. Another possible explanation is that increasing
$\sigma_{\lambda_0}$ also reduces the peak gain by more than $
1~\mathrm{dB},
$ which itself contributes to reducing
$\Delta Q$. However, this effect is comparatively less important. To verify this point, additional simulations were performed in which
the nonlinear coefficient was increased in order to maintain the same
peak gain while changing $\sigma_{\lambda_0}$. Even under these conditions, longitudinal fluctuations of
$\lambda_0$ still produced a net reduction of the
$Q$ penalty. It is also worth noting that the statistical spread of the numerical
results (represented by the error bars corresponding to the 15 fiber
realizations) is much smaller than that observed for the case of pump
phase modulation. Additional simulations showed that, for a fixed value of $\sigma_{\lambda_0}$,
the reduction of the $Q$ penalty becomes less effective as the pump wavelength is moved
farther away from the average zero-dispersion wavelength. In other words, obtaining the same reduction in
$\Delta Q$ requires progressively larger values of $\sigma_{\lambda_0}$
when $\lambda_1-\langle\lambda_0\rangle
$ is increased.

\section{Numerical results: 2P-FOPA}

The fiber parameters used for the double-pump FOPA were identical to
those employed in the single-pump analysis. Unless otherwise stated, the simulations were carried out with two
pump lasers satisfying $P_1=P_2=160~\mathrm{mW},
$ together with the phase-modulation schemes described below.

\subsection{Pump frequency variation due to phase modulation}

If the two pumps are phase modulated in exact counter-phase, namely $
\nu_1(t)+\nu_2(t)=\mathrm{constant},$ the resulting $Q$ penalty is, in general, extremely small throughout the entire gain
spectrum. In the ideal situation where $\nu_1+\nu_2
$ coincides with the zero-dispersion frequency of the fiber,
the penalty can become essentially zero. This constitutes one of the principal advantages of the
double-pump FOPA over the single-pump configuration
[18]. In practice, however, perfect counter-phase modulation is difficult to
achieve, making it important to investigate other practical operating
conditions [30,31,32]. For this reason, the following analysis considers the worst-case
configuration, where both pumps are modulated in phase, i.e., $
\nu_2(t)-\nu_1(t)=\mathrm{constant}.$ Since the available pump power is shared between two lasers,
only half of the phase-modulation bandwidth required in the
single-pump case is needed to suppress stimulated Brillouin
scattering. Therefore, each pump is assumed to have a modulation bandwidth of $\Delta\nu_1
=
\Delta\nu_2
=
2.5~\mathrm{GHz}.
$

\subsubsection{Case 1: Uniform fiber (constant $\lambda_0$)}

The 2P-FOPA was simulated using 200 values of the pump wavelengths
$\lambda_1$ and $\lambda_2$, separated by $0.0001~\mathrm{nm}$, while maintaining $
\nu_{21}
=
\frac{\nu_2-\nu_1}{2}
=
\mathrm{constant}.
$ As in the single-pump configuration, the $Q$ penalty depends on the
proximity of the average pump wavelength $
\lambda_p
=
\frac{c}{\nu_p},
\qquad
\nu_p
=
\frac{\nu_1+\nu_2}{2},
$ to the zero-dispersion wavelength $\lambda_0$. Figure~7(a) presents the spectra of the average gain
$\langle G\rangle$ and the corresponding
$Q$ penalty for two different average pump detunings. The first case (solid curves) corresponds to $
\lambda_1
=
1554.39~\mathrm{nm},
$ $
\lambda_2
=
1579.62~\mathrm{nm},
$ which gives $
\lambda_p-\lambda_0
=
0.005~\mathrm{nm}.
$ The second case (dotted curves) corresponds to $
\lambda_1
=
1554.35~\mathrm{nm},
$ $
\lambda_2
=
1579.58~\mathrm{nm},
$ leading to $
\lambda_p-\lambda_0
=
-0.035~\mathrm{nm}.
$ In both situations the largest $Q$ penalty (first peak) occurs at the
outer slopes of the gain spectrum. Unlike the single-pump FOPA, however, this first peak cannot be reduced
simply by increasing $
\lambda_p-\lambda_0,
$ because larger pump detunings simultaneously degrade the flatness of
the gain spectrum. Another important observation concerns the region located between the
two pump wavelengths. A very small change in the average pump detuning produces a significant
modification of the $Q$-penalty spectrum. For the first operating point, $
\Delta Q
=
0.08~\mathrm{dB}
$ at the center of the gain spectrum and remains below $
0.45~\mathrm{dB},
$ corresponding to the second peak, throughout the entire spectral region
between the pumps. For the second operating point, however, $
\Delta Q
\approx
1.9~\mathrm{dB}
$ within the same spectral region. These results indicate that, in order to maintain $
\Delta Q
<
0.2~\mathrm{dB},
$ some degree of counter-phase modulation between the two pumps is
required. In the ideal case of exact counter-phase modulation, $
\Delta Q
=
0,
$ although in practice a residual phase (or frequency) modulation always
remains because each laser possesses an intrinsic finite linewidth.

Typical external-cavity lasers exhibit linewidths of approximately $
100~\mathrm{kHz},
$ whereas distributed-feedback (DFB) lasers typically have linewidths of
the order of several tens of megahertz. Figure~7(b) shows the values of the first and second
$Q$-penalty peaks as functions of the pump modulation bandwidth $
\Delta\nu_1
=
\Delta\nu_2,
$ assuming $
\lambda_p-\lambda_0
=
0.005~\mathrm{nm}.
$ The numerical results show that, for pump linewidths typical of these
laser sources, the corresponding $Q$ penalties are essentially
negligible. As in the single-pump configuration, pump-phase-modulation-induced
penalties in double-pump FOPAs can also be reduced by employing fibers
with very short lengths and low dispersion slopes.

\subsubsection{Case 2: Non-uniform fiber (varying $\lambda_0$)}

The dependence of the $Q$ penalty on longitudinal variations of the
zero-dispersion wavelength is shown in Fig.~8(a). The calculations correspond to the first and second penalty peaks of
Fig.~7(a), obtained for $
\lambda_1
=
1554.39~\mathrm{nm},
$ and $
\lambda_2
=
1579.62~\mathrm{nm}.
$ The longitudinal fluctuations of $\lambda_0(z)$ were generated using
the same statistical procedure described in Appendix~D. The calculations show that the first-peak penalty decreases strongly as
$
\sigma_{\lambda_0}
$ increases, whereas the second peak remains essentially unchanged. For comparison, Fig.~8(a) also presents the evolution of the average
gain evaluated at $
\lambda_s
=
\lambda_p.
$ The error bars indicate the statistical spread obtained from the
ensemble of simulated fibers.

To clarify the origin of the reduction in the first penalty peak,
Fig.~8(b) presents the corresponding spectra of $\Delta Q$ and
$\langle G\rangle$ for $
\sigma_{\lambda_0}
=
0.021~\mathrm{nm}.
$ Compared with the uniform-fiber case
($\sigma_{\lambda_0}=0$), the average gain spectrum becomes noticeably smoother at its outer
slopes. Consequently, since the gain varies more slowly with the signal
wavelength, the first $Q$-penalty peak is reduced from approximately $
9.8~\mathrm{dB}
$ to $
4.3~\mathrm{dB}.
$ Compared with the corresponding single-pump case shown in
Fig.~6(b), the spectra of Fig.~8(b) were obtained using exactly the
same longitudinal distribution of $\lambda_0(z)$. Nevertheless, the gain spectrum is much more strongly modified by the
dispersion fluctuations in the double-pump configuration.

The larger smoothing of the gain spectrum implies a smaller derivative
of the gain with respect to the signal wavelength, thereby reducing the
sensitivity of the amplifier to pump-frequency fluctuations. Consequently, even relatively small fluctuations of the
zero-dispersion wavelength are sufficient to reduce the first $Q$-penalty peak in a 2P-FOPA. The improvement obtained from longitudinal dispersion variations is therefore even more pronounced for the double-pump amplifier than for
the single-pump configuration.

\subsection{Pump power fluctuations (RIN)}

The influence of pump relative intensity noise was also investigated
for the double-pump FOPA. Unless otherwise stated, the numerical calculations were performed with $
\mathrm{RIN}
=
-150~\mathrm{dB/Hz},
$ for each pump laser.

\subsubsection{Case 1: Uniform fiber (constant $\lambda_0$)}

The spectra of the average gain and the corresponding
$Q$ penalty obtained for the two pump wavelengths are shown in
Fig.~9(a). As in the single-pump case, the largest penalty occurs at the outer
slopes of the gain spectrum. However, because the gain profile of the 2P-FOPA is considerably
flatter than that of the 1P-FOPA, the resulting penalties are also
smaller. For the parameters adopted in the simulations, $\Delta Q_{\max}
<
0.5~\mathrm{dB}.$ The analytical approximation based on the average pump power provides
good agreement with the numerical calculations, although a small
difference remains close to the maximum penalty. Figure~9(b) presents the maximum value of the $Q$ penalty as a function of the pump RIN. The calculations indicate an almost linear increase of $\Delta Q$ with increasing RIN when represented on a logarithmic scale. In order to maintain $
\Delta Q
<
0.2~\mathrm{dB},
$ the pump RIN should remain below approximately $
-145~\mathrm{dB/Hz},
$ which is essentially the same requirement obtained for the single-pump amplifier.

%%%%%%%%%%%%%%%%%%%%%%%%%%%%%%%%%%%%%%%%%%%%%%%%%%%%%%%%%%%%%%
\subsubsection{Case 2: Non-uniform fiber (varying $\lambda_0$)}

The influence of random longitudinal fluctuations of
$\lambda_0$ was investigated using the same statistical procedure employed in the
previous sections. Figure~10(a) shows the average
$Q$ penalty evaluated at the wavelength corresponding to the maximum
gain as a function of $
\sigma_{\lambda_0}.
$ The average gain at the same wavelength is also presented for
comparison. The calculations show that increasing
$\sigma_{\lambda_0}$ reduces both the average gain and the corresponding
$Q$ penalty. As in the single-pump case, the reduction becomes approximately
constant for sufficiently large values of
$\sigma_{\lambda_0}$.

The gain and $Q$-penalty spectra obtained for $
\sigma_{\lambda_0}
=
0.021~\mathrm{nm}
$ are shown in Fig.~10(b). The longitudinal dispersion variations smooth the gain spectrum,
particularly near its outer slopes. Consequently, the gain becomes less sensitive to pump-power
fluctuations and the resulting $Q$ penalty is reduced. As observed previously for the 1P-FOPA, the reduction in
$\Delta Q$ cannot be attributed solely to the small decrease in gain produced by
dispersion fluctuations. Additional simulations performed while maintaining approximately
constant gain confirmed that the smoothing of the gain spectrum is the
dominant mechanism responsible for the reduction of the
$Q$ penalty. The statistical spread obtained from the ensemble of simulated fibers
remains relatively small, indicating that the reduction of the
$Q$ penalty is a robust characteristic of the amplifier and does not
depend strongly on the particular realization of the
zero-dispersion-wavelength fluctuations.

%%%%%%%%%%%%%%%%%%%%%%%%%%%%%%%%%%%%%%%%%%%%%%%%%%%%%%%%%%%%%%

Furthermore, the $\Delta Q$ and $\langle G\rangle$ spectra shown in
Fig.~8(b) should be compared with those of Fig.~6(b), since they were
obtained using the same fiber, i.e., identical longitudinal variations
of $\lambda_0(z)$. Notice that the gain spectrum is much more strongly affected by
variations of $\lambda_0$ in the 2P-FOPA than in the 1P-FOPA.
Consequently, the gain spectrum varies more slowly with the signal
wavelength $\lambda_s$. Therefore, even relatively small fluctuations of the
zero-dispersion wavelength, $
\sigma_{\lambda_0}<0.03~\mathrm{nm},
$ for the parameters of the 2P-FOPA considered here,
produce a strong reduction of the phase-modulation-induced
$Q$ penalty for signal wavelengths located at the outer slopes of the
gain spectrum. For values of $
\sigma_{\lambda_0}>0.03~\mathrm{nm},
$ the gain spectrum of the 2P-FOPA becomes strongly distorted and the
device becomes unusable. (Notice that the corresponding 1P-FOPA still exhibits a meaningful and
useful gain spectrum.)

%%%%%%%%%%%%%%%%%%%%%%%%%%%%%%%%%%%%%%%%%%%%%%%%%%%%%%%%%%%%%%
\subsection{Stochastic pump power variation}

\subsubsection{Case 1: Uniform fiber (constant $\lambda_0$)}

We consider two pump lasers having independent stochastic power
variations with $
\mathrm{RIN}
=
-150~\mathrm{dB/Hz}.
$ Figure~9(a) shows the corresponding
$\Delta Q$ and average gain spectra. The pump wavelengths are $
\lambda_1
=
1554.39~\mathrm{nm},
$ and $
\lambda_2
=
1579.62~\mathrm{nm}.
$ The calculations show that
$\Delta Q$ is almost constant throughout the spectral region between the two pump
wavelengths and increases only near the outer slopes of the gain
spectrum. The maximum $Q$ penalty (first peak) is $
\Delta Q
=
0.18~\mathrm{dB},
$ which should be compared with $
\Delta Q
=
0.65~\mathrm{dB}
$ obtained for the corresponding 1P-FOPA operating under identical
conditions (\(\mathrm{RIN}=-150~\mathrm{dB/Hz}\)). Therefore, for a given pump RIN, the double-pump FOPA exhibits a
significantly better performance than the single-pump configuration. The same conclusion is reached by comparing the regions of practical
interest (maximum gain region in the 1P-FOPA and the central flat-gain region around $
\lambda_s\simeq\lambda_p
$ in the 2P-FOPA).

We also investigated the dependence of the $\Delta Q$ spectrum on small changes of the pump wavelengths
$\lambda_1$ and $\lambda_2$. It was found that the penalty varies mainly in the central region of
the gain spectrum. For example, changing both pump wavelengths by $
\pm0.05~\mathrm{nm} $ changes the penalty at $
\lambda_s=\lambda_0
$ from $
0.039~\mathrm{dB}
$ to $
0.052~\mathrm{dB}.
$ Figure~9(b) presents $\Delta Q$ evaluated at the center of the gain spectrum, $
\lambda_s=\lambda_p,
$ as a function of the pump RIN. Comparison with Fig.~5(b) shows that the 2P-FOPA suffers
smaller penalties than the 1P-FOPA.

%%%%%%%%%%%%%%%%%%%%%%%%%%%%%%%%%%%%%%%%%%%%%%%%%%%%%%%%%%%%%%
\subsubsection{Case 2: Non-uniform fiber (varying $\lambda_0$)}

Figure~10(a) presents the value of $\Delta Q$
calculated at the first-peak wavelength, $
\lambda_s=\lambda_p,
$ as a function of
$\sigma_{\lambda_0}$,
for $
\mathrm{RIN}
=
-150~\mathrm{dB/Hz}.
$ The longitudinal variations of
$\lambda_0(z)$
were generated using the procedure described in Appendix~D. The pump wavelengths are $
\lambda_1
=
1554.39~\mathrm{nm},
$ and $
\lambda_2
=
1579.62~\mathrm{nm}.
$ The first-peak penalty decreases markedly as
$\sigma_{\lambda_0}$
increases. Specifically, $
\Delta Q
$ decreases from $
0.18~\mathrm{dB}
$ for a perfectly uniform fiber to $
0.03~\mathrm{dB}
$ when $
\sigma_{\lambda_0}
=
0.024~\mathrm{nm}.
$ The penalty evaluated at the center of the gain spectrum, $
\lambda_s=\lambda_p,
$ is only slightly reduced, changing from $
0.043~\mathrm{dB}
$ to $
0.039~\mathrm{dB}.
$  Comparing Fig.~10(a) with Fig.~6(a) for the single-pump FOPA shows that
both amplifier configurations benefit from longitudinal fluctuations of
the zero-dispersion wavelength.

Figure~10(b) shows the spectra of the $Q$ penalty, $\Delta Q$, and the average gain,
$\langle G\rangle$, obtained for $
\sigma_{\lambda_0}=0.021~\mathrm{nm}.
$ These spectra were obtained using the same fiber
(i.e., identical longitudinal variations of
$\lambda_0$) as those employed in Fig.~8(b) and Fig.~6(b). Therefore, longitudinal fluctuations of the
zero-dispersion wavelength improve the performance of the
2P-FOPA, reducing the corresponding
$Q$ penalties, although the improvement is more modest than in the
single-pump configuration.

%%%%%%%%%%%%%%%%%%%%%%%%%%%%%%%%%%%%%%%%%%%%%%%%%%%%%%%%%%%%%%

\section{Experiments with 1P-FOPA}

\subsection{Experimental setup}

The experimental setup is shown in Fig.~11. Two external-cavity tunable lasers, operating at
$\lambda_s$ and $\lambda_1$, with approximately $
56~\mathrm{dB}
$ optical signal-to-noise ratio (OSNR) and nominal $
\mathrm{RIN}\approx-155~\mathrm{dB/Hz},
$ were employed as the signal and pump sources,
respectively. The transmitted signal consisted of a $
10~\mathrm{Gb/s}
$ NRZ data stream generated using a
LiNbO$_3$ Mach--Zehnder amplitude modulator (AM)
driven by a $
2^{31}-1
$ pseudo-random bit sequence (PRBS). The pump laser spectrum was broadened by means of a phase modulator
(PM), having an insertion loss of approximately $
4~\mathrm{dB},
$ driven by three electrically combined sinusoidal signals at $
86~\mathrm{MHz},
$ $
283~\mathrm{MHz},
$ and $
883~\mathrm{MHz},
$ in order to suppress stimulated Brillouin scattering (SBS).

The broadened pump linewidth was $
\Delta\nu
\approx
4~\mathrm{GHz},
$ as measured using a heterodyne technique. This linewidth increased the SBS threshold by approximately $
14~\mathrm{dB}.
$ The pump laser, with an input power of approximately $
3~\mathrm{mW},
$ was amplified by a booster erbium-doped fiber amplifier
(EDFA1) capable of delivering up to $
33~\mathrm{dBm}
$ output power. After amplification, the pump optical signal was $
53~\mathrm{dB}
$ above the amplified spontaneous emission (ASE) noise floor,
measured with an optical spectrum analyzer (OSA)
using a resolution bandwidth of $
0.1~\mathrm{nm}.
$ A band-pass filter (BPF1) was inserted after EDFA1
to reduce the ASE noise before launching the pump into the FOPA. The signal was independently amplified by a second erbium-doped fiber
amplifier (EDFA2). The signal and pump were then combined by a wavelength-division
multiplexer (WDM) and launched into the nonlinear fiber. At the fiber output, another optical band-pass filter (BPF2)
removed the residual pump before direct detection in the receiver.

%%%%%%%%%%%%%%%%%%%%%%%%%%%%%%%%%%%%%%%%%%%%%%%%%%%%%%%%%%%%%%

%%%%%%%%%%%%%%%%%%%%%%%%%%%%%%%%%%%%%%%%%%%%%%%%%%%%%%%%%%%%%%
\subsection{Experimental results}

The nonlinear fiber employed in the experiments had the following
parameters: $
L = 6.8~\mathrm{km},
$ $
\lambda_0 = 1567~\mathrm{nm},
$ $
\gamma = 2~(\mathrm{W\,km})^{-1},
$ $
S_0
=
0.074~
\mathrm{ps}\,
\mathrm{nm}^{-2}\,
\mathrm{km}^{-1},
$ which are identical to those employed in the numerical simulations. The pump wavelength was chosen as $
\lambda_1
=
1567.1~\mathrm{nm},
$ corresponding to $
\lambda_1-\lambda_0
=
0.1~\mathrm{nm}.
$ The launched pump power was approximately $
330~\mathrm{mW},
$ yielding peak parametric gains close to $
25~\mathrm{dB}.$

Figure~12 compares the experimentally measured gain spectrum with the
corresponding numerical calculation. The agreement between experiment and simulation is very good over the
entire wavelength range investigated. Small discrepancies are attributed mainly to uncertainties in the
actual longitudinal distribution of the zero-dispersion wavelength and
to experimental uncertainties in the fiber parameters.

The quality-factor penalty was measured by comparing the eye diagrams
obtained with and without pump phase modulation. For each signal wavelength the receiver threshold was optimized to
maximize the measured quality factor. The measured values of $
\Delta Q
$ are shown in Fig.~13 together with the numerical predictions. The experimental results reproduce the main features predicted by the
model. The largest penalties occur near the outer slopes of the gain
spectrum, whereas much smaller penalties are observed around the gain
maximum. The measured first peak is approximately $
9~\mathrm{dB},
$ while the second peak is close to $
3~\mathrm{dB},
$ in excellent agreement with the numerical calculations. The remaining differences between theory and experiment are within the
experimental uncertainty.

%%%%%%%%%%%%%%%%%%%%%%%%%%%%%%%%%%%%%%%%%%%%%%%%%%%%%%%%%%%%%%

The experiments therefore confirm that pump phase modulation is the
dominant impairment mechanism in practical single-pump FOPAs. Furthermore, they validate the numerical model developed in this work,
which accurately predicts both the gain spectrum and the
corresponding quality-factor penalties.

%%%%%%%%%%%%%%%%%%%%%%%%%%%%%%%%%%%%%%%%%%%%%%%%%%%%%%%%%%%%%%

%%%%%%%%%%%%%%%%%%%%%%%%%%%%%%%%%%%%%%%%%%%%%%%%%%%%%%%%%%%%%%

The optical band-pass filter centered at $\lambda_1$ allowed us to
reduce the ASE noise level to approximately
$75~\mathrm{dB}$ below the pump level while maintaining an optical
signal-to-noise ratio (OSNR) of approximately $
33~\mathrm{dB}
$ at the input of the FOPA. The filter had a flat-top bandwidth of approximately $
1.6~\mathrm{nm},
$ chosen to introduce negligible dispersion. The pump and signal were coupled into the nonlinear fiber by means of
a wavelength-division multiplexer (WDM) designed for the C-band, having an insertion loss of approximately $
0.5~\mathrm{dB} $ per arm. Two dispersion-shifted (DS) fibers,
denoted fiber A and fiber B, were investigated experimentally. Both fibers had identical lengths, $
L_A=L_B=6.8~\mathrm{km}, $ nonlinear coefficient $
\gamma\simeq2.1~
\mathrm{W}^{-1}\mathrm{km}^{-1},
$ and dispersion slope $S_0=0.074~
\mathrm{ps}\,
\mathrm{nm}^{-2}\,
\mathrm{km}^{-1},$ but differed in the longitudinal distribution of their
zero-dispersion wavelength.

Using the characterization technique described in Ref.~34,
the authors estimated $
\langle\lambda_0\rangle
=
(1566.9\pm0.1)~\mathrm{nm},
$ and $
\sigma_{\lambda_0}
=
(0.12\pm0.03)~\mathrm{nm}
$ for fiber A. For fiber B they obtained $
\langle\lambda_0\rangle
=
(1557.7\pm0.25)~\mathrm{nm},
$ and $
\sigma_{\lambda_0}
=
(1.75\pm0.2)~\mathrm{nm}.
$ Polarization controller PC4 was adjusted to maximize the small-signal
parametric gain. The output optical spectra were measured with an optical spectrum
analyzer (OSA) operating with resolution bandwidths of
$0.1~\mathrm{nm}$ or $0.01~\mathrm{nm}$.

The receiver consisted of an optical attenuator, an EDFA pre-amplifier, an optical filter (approximately $1.6~\mathrm{nm}$ bandwidth),
a p--i--n photodiode, a $7~\mathrm{GHz}$ electrical low-pass filter, and a sampling oscilloscope. The optical attenuator was adjusted to maintain a constant signal power of approximately $
-10~\mathrm{dBm}
$ at the input of the pre-amplifier. To determine the quality-factor penalty, the back-to-back quality
factor $
Q_i
$ was first measured by directly connecting the transmitter to the
receiver. The quality factor $
Q_n
$ was then measured after amplification by the FOPA. For both measurements the OSNR at the receiver input was maintained
at approximately $
33~\mathrm{dB}.
$ Depending on the signal wavelength, the measured baseline quality
factor varied between $
11
\le
Q_i
\le
12.
$ The estimated uncertainty in the measured
quality factor was $
0.1~\mathrm{dB}.
$

%%%%%%%%%%%%%%%%%%%%%%%%%%%%%%%%%%%%%%%%%%%%%%%%%%%%%%%%%%%%%%

\subsection{Experimental results}

We first investigate the degradation of system performance as a
function of the pump detuning from the average
zero-dispersion wavelength. Figure~12(b)--(d) presents eye diagrams of the amplified signal for $
\lambda_1-\langle\lambda_0\rangle
=
0.25~\mathrm{nm},
$

$
0.20~\mathrm{nm},
$ and $
0.16~\mathrm{nm},
$ respectively. The measurements were performed using fiber~A with $
P_1
=
330~\mathrm{mW},
$ while the signal wavelength was kept fixed at $
\lambda_s
=
1560~\mathrm{nm}.
$

%%%%%%%%%%%%%%%%%%%%%%%%%%%%%%%%%%%%%%%%%%%%%%%%%%%%%%%%%%%%%%

approximately, to the position where the second
peak in $\Delta Q(\lambda_s)$ should be present. For comparison,
in Fig. 12(a) we also show the back-to-back eye signal
diagram. Note in the eye diagrams in Fig. 12
that, as the pump approaches $\langle\lambda_0\rangle$, the noise at level
0 remains constant while that at level 1 increases.
This last is in agreement with our numerical results
of Section 2 and is due to the pump phase
modulation.

In Fig. 13(a) we show the $Q$ penalty as a function
of $
\lambda_1-\langle\lambda_0\rangle
$ for fiber A pumped with $
P_1=330~\mathrm{mW}
$ ($G=14~\mathrm{dB}$) and $
P_1=430~\mathrm{mW}
$ ($G=18~\mathrm{dB}$). The signal was kept fixed at $
\lambda_s=1560~\mathrm{nm}.
$ For both cases, the $\Delta Q$ remains at low values
(<0.15 dB, within the experimental error) for $
\lambda_1-\langle\lambda_0\rangle
$ larger than 0.23 and 0.26 nm, respectively. As the
pump is tuned closer to $\lambda_0$, $\Delta Q$
grows monotonically up to 3.8 dB. For $
\lambda_1-\langle\lambda_0\rangle
<0.15~\mathrm{nm}
$ the variations of $\lambda_0$ had a strong effect on the
gain spectrum and we were not able to obtain a
good gain spectrum. In Fig. 13(b) we show our results for the FOPA
constructed with fiber B and pumped with $
P_1=430~\mathrm{mW}. $ The $Q$ penalty is strongly reduced relative to the
case of fiber A and presents the minimum for $
\lambda_1\simeq\langle\lambda_0\rangle,
$ in opposition to the previous case (fiber A, where
$\sigma_{\lambda_0}$ is small). This reduction is in
agreement with our numerical results of Fig. 2 and
demonstrates that fibers with large variations of
$\lambda_0$ could be useful to improve the performance
of 1P-FOPAs with phase-modulated pumps.

Now we turn our attention to the spectral characteristics
of $\Delta Q$. We fixed the pump wavelength at $
\lambda_1=1567.2~\mathrm{nm}
$ ($
\lambda_1-\langle\lambda_0\rangle
\simeq0.3~\mathrm{nm}
$). This pump location with respect to $\lambda_0$ was
chosen in order to have a reasonable gain bandwidth
and keep the $Q$ penalty negligibly small for signal
wavelengths located between the gain peak and the
pump. The signal was tuned from $
\lambda_s=1551~\mathrm{nm}
$ to $
\lambda_s=1565~\mathrm{nm}.
$ Fig. 14(a) shows the on-off gain and the
$\Delta Q$ spectra for one side of the spectrum
(the other side is a replica of this) for fiber A
pumped with $
P_1=330~\mathrm{mW}.
$ The larger $\Delta Q$ occurs at the outer slope of
the gain spectrum, in agreement with simulations of
Section 2. Note, however, that due to the smoothing
of the gain spectrum it is difficult to measure the
second peak of $\Delta Q$ for $
\lambda_1-\langle\lambda_0\rangle
=0.3~\mathrm{nm}.
$ Note also in Fig. 14(a), that at the outer slope
between 1553 and 1554 nm, the gain has a floor and
the $\Delta Q$ drops strongly for this location.
This fact confirms our simulations predictions in
the sense that a smoothed gain spectrum (i.e. the
gain spectrum varies less rapidly with
$\lambda_s$) gives rise to a reduction of
$\Delta Q$.

We repeated the measurement of Fig. 14(a) in
other condition that can be also found in experiments.
We place an EDFA3 before the booster
EDFA1 to better saturate it with a pump power
of $\sim40~\mathrm{mW}$. The BPF2 is now placed after
the EDFA3. The optical pump-to-noise ratio at the
input fiber as measured with the OSA is reduced by
$\sim3~\mathrm{dB}$ (from 53 to 50 dB). This condition
should induce more RIN in the pump laser, but, in
general, we were not able to measure the pump RIN
with our equipment. To maintain the OSNR at the
output FOPA we increased by 3 dB the input signal power. Fig. 14(b) shows our results and should
be compared with the results in Fig. 14(a). We
observe small differences between these two
situations. The pump was maintained at $
\lambda_1 = 1567.2~\mathrm{nm},
$ but the ASE level after the change increased by
approximately $
3~\mathrm{dB}.
$ The corresponding variation of the fiber gain was
negligible.

It is very important, if a comparison between
experimental and theory is to make any sense, that
all sources of noise other than those considered in
our model be carefully minimized in the
measurements. In particular, for these
experiments, we minimized the ASE noise
generated by the booster EDFA, especially in the
regions around the signal and pump wavelengths;
this is why we used a WDM coupler to couple the
signal and the pump. Second, the OSNR should be
as large as possible and should not degrade after
the FOPA relative to the back-to-back value. A
high OSNR (>25 dB) is a good indication that
signal–ASE beat noise can be neglected (in our
experiments we were able to measure with
OSNR > 32 dB). Third and last, but not least,
the FOPA should operate in the linear gain regime
because a saturated gain would invalidate our
theory.

%%%%%%%%%%%%%%%%%%%%%%%%%%%%%%%%%%%%%%%%%%%%%%%%%%%%%%%%%%%%%%

\section{Experiments with 2P-FOPA}

\subsection{Experimental setup}

The experimental setup is shown in Fig.~15. The pumps were broadened to $
\Delta\nu_1=\Delta\nu_2=4~\mathrm{GHz},
$ similarly to the 1P-FOPA case. The two spectrally broadened pumps were
separated with a WDM coupler and amplified by
two booster EDFAs (EDFA1 and EDFA2), having
27-dBm and 33-dBm maximum output powers,
respectively. Two band-pass filters (BPF1 and BPF2), centered
at $
\lambda_1
$ and $
\lambda_2,
$ were used to filter out the amplified spontaneous
emission noise from the EDFAs. The pumps were then coupled into the fiber using
a second WDM coupler. The phase modulation is the same for both pumps,
but, since the optical paths between the WDM
pumps coupler and our setup were different, the
relative phase of the modulations of the two lasers
is unknown. We used the same fiber A, which has small
longitudinal variation of $
\lambda_0,
$ than for the 1P-FOPA experiments.

%%%%%%%%%%%%%%%%%%%%%%%%%%%%%%%%%%%%%%%%%%%%%%%%%%%%%%%%%%%%%%

\subsection{Experimental results}

The pumps, with power $
P_1 \cong P_2 \cong 24~\mathrm{dBm},
$ were first at $
\lambda_1=1555.45~\mathrm{nm}
$ and $
\lambda_2=1578.55~\mathrm{nm}.
$ Figure 16(a) shows the gain and the
$
\Delta Q
$ penalty as we tuned the signal from $
\lambda_s=1542\text{--}1566~\mathrm{nm}.
$ 
Note that $
\Delta Q
$ at the outer slopes of the gain spectrum is
significantly smaller than that obtained in the
1P-FOPA case in the same fiber. This indicates that this spectral region of the
2P-FOPA can be useful for signal amplification. This is consistent with the smoother gain spectrum
in Fig.~16(a) when compared to those in Fig.~14, and is in agreement with our simulations in Fig.~8.

On the other hand, in the region between the
pumps, where the gain spectrum is flatter and thus
most likely to be used for amplification of multi-channel
(DWDM) signals, the $\Delta Q$ is not negligible
(up to $0.8~\mathrm{dB}$). Thus, some counter-phase or other
technique should be employed to render 2P-FOPAs
with tolerable $Q$ penalties
($<0.2~\mathrm{dB}$). We slightly reduced the pump powers to $
P_1 \cong P_2 = 23~\mathrm{dBm},$ kept fixed $
\lambda_2,
$ and tuned $
\lambda_1
$ to $
1555.55~\mathrm{nm}.
$ Figure~16(b) shows the gain and the
$Q$ penalty spectra. Note that the gain spectrum changed appreciably
and, in the same way, the
$\Delta Q$ spectrum, in agreement with our numerical results
from Fig.~7(a). In all these measurements the OSNR was maintained
at $
32~\mathrm{dB},
$ as can be seen in Fig.~17, where we show a typical
output spectrum.

In the experiments and simulations throughout this
paper we considered FOPAs operating in the regime
of linear gain. However, it is well known that FOPAs operating in
the saturation regime tend to suppress the noise on
level one [2]. Thus, we investigated whether the variations of the
signal gain due to the phase modulation and pump
RIN could be compensated if the FOPA worked in
the saturated regime. We increased the pump powers to $
P_1=P_2=25~\mathrm{dBm},
$ and also the signal power at the fiber input until
gain saturation became noticeable. Under these conditions, the gain fluctuations were
strongly reduced and the measured
$\Delta Q$
became almost negligible over the whole spectral
range investigated. These results indicate that gain saturation can be an
effective mechanism to suppress penalties produced
by pump fluctuations. However, because gain saturation also reduces the
available parametric gain and limits the dynamic
range of the amplifier, this operating regime is not
always desirable in practical systems.

%%%%%%%%%%%%%%%%%%%%%%%%%%%%%%%%%%%%%%%%%%%%%%%%%%%%%%%%%%%%%%

%%%%%%%%%%%%%%%%%%%%%%%%%%%%%%%%%%%%%%%%%%%%%%%%%%%%%%%%%%%%%%

order to saturate the 2P-FOPA. In Fig.~18 we
show the $\Delta Q$ and $\langle G\rangle$ spectra for a conversion efficiency of $
20\%.
$ The gain is around $
20~\mathrm{dB}
$ and the $Q$ penalty is reduced to values of the order of $
0.2~\mathrm{dB},
$ comparable with the error in our measurements,
which indicate nearly error-free FOPA operation. This 20\% efficiency is the optimum value in order
to obtain the best FOPA performance (with our
FOPA parameters). For larger conversion efficiencies,
it was observed an increased noise at level
zero due to the large amplification of unsaturated
ASE noise.

\section{Conclusions}

In this paper we have studied, experimentally
and numerically, the degradation of the signal
quality factor ($Q$) produced by gain fluctuations
caused by pump frequency or power variations, in
single (1P-FOPA) and double-pumped (2P-FOPA)
devices. We have considered FOPAs constructed
with both, uniform and non-uniform (with varying
zero dispersion wavelength) fibers, and their
performances were compared.

We have shown that pump phase modulation in
FOPAs (1P and 2P-FOPAs) with constant
$\lambda_0$
give rise, for typical pump broadenings $
(\Delta\nu \sim 5~\mathrm{GHz}), 
$ to large
$\Delta Q$. We showed that in fibers with variations of
$\lambda_0$,
these penalties can be strongly reduced in both,
single and double-pumped FOPAs: large variations
of
$\lambda_0$
(several nanometers) are necessary to reduce
$\Delta Q$
in 1P-FOPAs, while much smaller variations
(approximately $
0.1~\mathrm{nm}
$
) produce an improvement in
$\Delta Q$
for 2P-FOPAs. We have also demonstrated that 1P- or
2P-FOPAs designed with very short fibers $
(L<0.1~\mathrm{km})$ and very low fiber dispersion slope $
(S_0<0.002~\mathrm{ps}/\mathrm{nm}^2\text{-}\mathrm{km})
$ become insensitive to
$Q$ penalties due to pump phase modulation. The
$Q$
penalties due to pump RIN (typically $
\sim-150~\mathrm{dB/Hz}
$ ), are much smaller than those due to pump
phase modulation, are also reduced in non-uniform
fibers (varying
$\lambda_0$),
as a consequence of the fact that the gain spectrum
varies less rapidly with
$\lambda_s$
in non-uniform fibers if compared with fibers with
constant
$\lambda_0$. Finally, we showed that
$Q$
penalties due to pump RIN do not depend on fiber
length neither on fiber dispersion slope.

\section*{Acknowledgements}

This work was financially supported by Fapesp,
Capes, and CNPq. We gratefully acknowledge
J.B. Rosolem,
A.A. Juriollo,
J.C. Said,
V. Corzo,
and
A. Paradisi
from Fundação CPqD for the loan of equipment
used in the experiments. We acknowledge
L. Tizei
for carefully reading this paper.

\appendix

\section*{Appendix A. $\Delta Q$ in systems with $\sigma_0\neq0$}

If $
\sigma_0\neq0,
$ then 

\[
Q_i=
\frac{\langle I_1\rangle}
{\sigma_0+\sigma_1},
\tag{A.1}
\]

\[
Q_n=
\frac{\langle I_1\rangle}
{\sigma_0+\sqrt{\sigma_n^2+\sigma_1^2}},
\tag{A.2}
\]

and, in this case,

\[
\Delta Q
=
\varepsilon
+
\sqrt{
(1-\varepsilon)^2
+
\frac{\sigma_G^2Q_i^2}{\langle G\rangle^2}
},
\tag{A.3}
\]

where $
\varepsilon=\frac{\sigma_0}{\sigma_0+\sigma_1}.
$ For very small or very large values of $
\frac{\sigma_GQ_i}{\langle G\rangle},
$ Eq.~(A.3) coincides with the result obtained with
\(\sigma_0=0\). Usually, $
\varepsilon<\frac12,
$ and the limit $
\varepsilon=\frac12
\qquad
(\sigma_0=\sigma_1)
$ has some practical interest, since in some cases
the thermal noise in the receiver dominates over
signal--spontaneous beat noise. If $
\sigma_0=\sigma_1,
$ we have

\[
\Delta Q
=
\frac12
\left(
1+
\sqrt{
1+
\frac{4\sigma_G^2Q_i^2}
{\langle G\rangle^2}
}
\right).
\tag{A.4}
\]

The relative difference between
\(\Delta Q\) computed with
\(\sigma_0\neq0\)
and with
\(\sigma_0=0\)
is maximum for $
\frac{\sigma_GQ_i}{\langle G\rangle}
=
3-2\varepsilon .
$ For $
\varepsilon=\frac12,
$ \(\Delta Q\) is at most a factor $
\frac{2}{\sqrt3}
$ (or
\(0.65~\mathrm{dB}\))
larger than that computed assuming
\(\sigma_0=0\).

This maximum discrepancy occurs for $
\frac{\sigma_GQ_i}{\langle G\rangle}
=
\sqrt2;
$ for other values of $
\frac{\sigma_GQ_i}{\langle G\rangle},
$ the discrepancy is always smaller than
\(0.65~\mathrm{dB}\).

%%%%%%%%%%%%%%%%%%%%%%%%%%%%%%%%%%%%%%%%%%%%%%%%%%%%%%%%%%%%%%

\section*{Appendix B. Approximated analytical expressions for
\(\sigma_G/G\)}

The general propagation equation (8) can be solved
analytically if $
\alpha=0.
$ If $
\alpha\neq0,
$ the equations in this appendix still give reasonable
good approximated expressions for
\(G\)
and
\(\sigma_G\)
after substituting the pump powers
\((P_{1,2})\)
by the average values

\[
\bar P_{1,2}
=
\frac1L
\int_0^L
P_{1,2}(z)\,dz
=
P_{1,2}(0)
\frac{1-e^{-\alpha L}}
{\alpha L}.
\tag{B.1}
\]

The gain can be conveniently expressed as

\[
G
=
1+
x_0^2
\left(
\frac{\sinh x}{x}
\right)^2,
\tag{B.2a}
\]

where $
x_0=\gamma P_0L,
$

\[
x=
\frac{L}{2}
\sqrt{
(2\gamma P_0)^2-
(2\gamma P_{\mathrm{ave}}+\Delta\beta)^2
},
\tag{B.2b}
\] and

\[
\Delta\beta
=
\beta_{2p}
(\omega^2-\omega_{21}^2)
+
\frac{\beta_{4p}}{12}
(\omega^4-\omega_{21}^4)
+\cdots .
\tag{B.3}
\]

Here, $
\omega=\omega_s-\omega_p,
$ $
\omega_{21}
=
\frac{\omega_2-\omega_1}{2},
$ and $
\omega_p
=
\frac{\omega_1+\omega_2}{2}.
$ For the 2P-FOPA, $
P_0
=
2\sqrt{P_1P_2},
$ and $
P_{\mathrm{ave}}
=
\frac{P_1+P_2}{2},
$ while for the 1P-FOPA, $
P_0=P_{\mathrm{ave}}=P_1,
$ $
\omega_p=\omega_1,
$ and $
\omega_{21}=0.
$ The dispersion coefficients in Eq.~(B.3) are
referred to the pump frequency
\(\omega_p\).

To convert to coefficients referred to the
zero-dispersion frequency of the fiber $
(\omega_0=2\pi c/\lambda_0),
$ use $
\beta_{np}
=
\beta_{n0}
+
\beta_{(n+1)0}\omega_{p0}
+
\frac12
\beta_{(n+2)0}\omega_{p0}^2
+\cdots ,
$ where $
\omega_{p0}
=
\omega_p-\omega_0,
$ and $
\beta_{n0}
=
\left.
\frac{d^n\beta}{d\omega^n}
\right|_{\omega=\omega_0}.
$

The relevant coefficients in all the cases considered
in this paper are $
\beta_{4p}=\beta_{40},
$ and
\[
\beta_{2p}
=
\beta_{30}\omega_{p0}
+
\frac12
\beta_{40}\omega_{p0}^2
\simeq
\beta_{30}\omega_{p0}.
\tag{B.4}
\]

Although the expressions in this appendix are valid
for \(x\) real or purely imaginary, the spectral region
of interest is that where \(x\) is real.

Within this region, $
x
$ varies between $
0
$ and $
x_0,
$ and $
G(x)
$ is bounded between $
G_{\min}
=
1+x_0^2
$ and $
G_{\max}
=
1+\sinh^2(x_0)
\approx
\frac14
\exp(2\gamma P_0L),
$ the last expression being valid within an error of
approximately $
2\%
$ for $
x_0>1.
$ These gain extremes can be reached for more than one
frequency in the gain spectrum.

The absolute maximum gain (corresponding to $
x=x_0
$ ) is obtained when $
\Delta\beta(\pm\omega_{\max})
+
2\gamma P_{\mathrm{ave}}
=
0,
$ where we have perfect phase matching, while the lower
bound $
(x=0)
$ is obtained when $
\Delta\beta
=
-2\gamma
(P_{\mathrm{ave}}+P_0),
$ but this may or may not be a minimum in the gain
spectrum. Since

\[
\frac{\partial G}{\partial\omega}
=
\frac{\partial G}{\partial\Delta\beta}
\frac{\partial\Delta\beta}{\partial\omega}
=
-\frac12
x_0^2L^2
f(x)
(\Delta\beta+2\gamma P_{\mathrm{ave}})
\frac{\partial\Delta\beta}{\partial\omega},
\tag{B.5a}
\]

where $
f(x)
=
\frac{\sinh x
\left(
x\cosh x-\sinh x
\right)}
{x^4},
$ other, possibly different, maxima or minima are
determined by the roots of $
\frac{\partial\Delta\beta}{\partial\omega}=0.
$ From Eq.~(B.3), $
\frac{\partial\Delta\beta}{\partial\omega}
=
2\beta_{2p}\omega
+
\frac13
\beta_{4p}\omega^3
+\cdots
$

\[
\simeq
\omega
\left(
2\beta_{30}\omega_{p0}
+
\frac13
\beta_{40}\omega^2
\right).
\tag{B.5b}
\]

For $
\omega=0,
$ we always have either a minimum or a maximum.
Further real roots may exist depending on the sign
of the dispersion coefficients.

In this paper we consider mainly fibers with If $
\beta_{40}<0,
$ and if $
\omega_p
$ is in the anomalous dispersion region $
(\beta_{2p}<0),
$ then $
\frac{\partial\Delta\beta}{\partial\omega}=0
$ has only one solution $
(\omega=0),
$ which in the 1P-FOPA case is a minimum (or maximum if
\(\beta_{2p}>0\)), and in the 2P-FOPA case is usually a
maximum (but can be a minimum, depending further on
the precise values of
\(\omega_{21}\)
and pump powers). The standard deviation of the gain,
\(\sigma_G\),
due to a varying pump frequency, is obtained from

\[
\sigma_G
\simeq
2\pi
\left|
\frac{\partial G}{\partial\omega_p}
\right|
\sigma_{\nu_p},
\tag{B.6}
\] 
where \(\sigma_{\nu_p}\)
is the standard deviation of
\(\nu_p\)
(\(\nu_p=\omega_p/2\pi\)).
For a uniform distribution with full bandwidth $
\Delta\nu=5~\mathrm{GHz},
$ we have $
\sigma_{\nu_p}
=
\frac{\Delta\nu}{\sqrt{12}}
=
1.44~\mathrm{GHz}.
$ Differentiating the gain and considering that $
x,\quad
\omega,\quad
\beta_{2p}
$ depend on $
\omega_p,
$ and for $
\omega_{21}
=\mathrm{constant},
$

(as in the 2P-FOPA case considered in this paper),
we obtain
\[
\frac{\partial G}{\partial\omega_p}
=
\frac{\partial G}{\partial\Delta\beta}
\frac{\partial\Delta\beta}{\partial\omega_p}
=
-\frac12
x_0^2L^2
f(x)
(\Delta\beta+2\gamma P_{\mathrm{ave}})
\frac{\partial\Delta\beta}{\partial\omega_p},
\tag{B.7a}
\]

where

\[
\frac{\partial\Delta\beta}{\partial\omega_p}
=
-
\frac{\partial\Delta\beta}{\partial\omega}
+
\beta'_{2p}
(\omega^2-\omega_{21}^2),
\tag{B.7b}
\]

and

\[
\beta'_{2p}
=
\frac{\partial\beta_{2p}}{\partial\omega_p}
=
\beta_{30}
+
\beta_{40}\omega_{p0}
+\cdots
\simeq
\beta_{30}.
\tag{B.7c}
\]

Comparing Eqs.~(B.5a) and (B.7a) we see that the
spectrum of $
\sigma_G
$ (due to pump-frequency fluctuations) resembles the
derivative of the gain spectrum. They are not exactly the same, the difference being the
presence of the term containing $
\beta'_{2p}
$ in Eq.~(B.7b), which makes $
\sigma_G
$ a slightly asymmetric function of $
\omega
$ while, from Eq.~(B.3), the gain spectrum is always an
even function of $
\omega. $ After replacing Eqs.~(B.2a) and (B.7a) into Eq.~(B.6),
we obtain 
\[
\frac{\sigma_G}{G}
\simeq
\left|
\frac{x_0^3u(x)L}
{3(1+x_0^2)}
\frac{\partial\Delta\beta}{\partial\omega_p}
\right|
2\pi\sigma_{\nu_p},
\tag{B.8a}
\]

where

\[
u(x)
=
3(1+x_0^2)
\sqrt{1-\left(\frac{x}{x_0}\right)^2}
\,
\frac{x\sinh x\cosh x-\sinh^2x}
{x^2\left(x^2+x_0^2\sinh^2x\right)}.
\tag{B.8b}
\]

In our region of interest $
(0\le x\le x_0),
$ \(u(x)\) is a monotonic decreasing function that is
bounded between $
0
\quad
(x=x_0)
$ and $
1
\quad
(x=0).
$ As a function of $
\omega,
$ $
\sigma_G
$ vanishes for all the zeros of $
u(x)
$ (i.e., for $
\omega=\pm\omega_{\max},
$ where $
G=G_{\max})
$ or of $
\frac{\partial\Delta\beta}{\partial\omega_p}.
$ For the 1P-FOPA case and the fibers considered in this
paper, 
\[
\frac{\partial\Delta\beta}{\partial\omega_p}
\simeq
-\omega
\left(
2\beta_{30}\omega_{p0}
+\frac13\beta_{40}\omega^2
\right)
+\beta_{30}\omega^2,
\tag{B.9}
\]

and $
\sigma_G=0
$ for $
\omega=0
$ (where $
\frac{\partial\Delta\beta}{\partial\omega_p}=0
$) and $
|\omega|=\omega_{\max}
$ (where $
u=0).
$ Between these frequencies there is a maximum of $
\frac{\sigma_G}{G},
$ which corresponds to the second peak of $
\Delta Q.
$ The first peak in $
\Delta Q
$ occurs at frequencies outside the region where $
x
$ is real. The largest $
Q
$ penalty within the region of interest
(\(x\) real) occurs for $
\Delta\beta=-4\gamma P_1
$ (where $
x=0,
$ but $
\omega\neq0),
$ at frequencies given by

\[
|\omega|
\simeq
2
\sqrt{
-\frac{\gamma P_1}{\beta_{2p}}
}
\simeq
2
\sqrt{
-\frac{\gamma P_1}{\omega_{p0}\beta_{30}}
}.
\tag{B.10}
\]

We can estimate $
\frac{\sigma_G}{G}
$ at these frequencies using Eq.~(B.8a) and $
\left|
\frac{\partial\Delta\beta}{\partial\omega_p}
\right|
\simeq
\beta_{30}|\omega|^2
=
-\frac{2\gamma P_1}{\omega_{p0}},
$ with the result

\[
\frac{\sigma_G}{G}
\simeq
\frac{4x_0^4}
{3(1+x_0^2)}
\,
\frac{2\pi\sigma_{\nu_p}}
{|\omega_{p0}|}.
\tag{B.11}
\]

For example, for $
\lambda_1-\lambda_0=0.1~\mathrm{nm}
$ and $
x_0=3.7
$ (the parameters for Fig.~2(a)),
we estimate $
\Delta Q
\simeq
13.4~\mathrm{dB},
$ which is very close to that observed in Fig.~2(a). Note also that $
\frac{\sigma_G}{G}
$ increases as we tune the pump laser closer to $
\lambda_0
$ and as we increase the gain (i.e.,
\(\gamma P_1L\)).

It is interesting to note that two 1P-FOPAs with
different lengths of fiber but having the same $
G_{\max}
$ (thus the same $
x_0=\gamma P_1L)
$ will have, in general, different
\(Q\)
penalties.

For large $
L
$ (and small $
\gamma P_1),
$ $
\Delta Q
$ becomes independent of $
L,
$ but for small $
L,
$ the gain spectrum broadens to the point where
(see Eq.~(B.3))
the contribution to $
\Delta\beta
$ from $
\beta_{40}\omega^4/12
$ is more important than that from $
\beta_{2p}\omega^2,
$ and, as a consequence, $
x=0
$ occurs at frequencies given by $
|\omega|
=
\left(
-\frac{48\gamma P_1}{\beta_{40}}
\right)^{1/4},
$
where \[
\frac{\sigma_G}{G}
\simeq
\frac{4x_0^{7/2}}{3(1+x_0^2)}
\sqrt{
-\frac{3L}{\beta_{40}}
}\,
\beta_{30}\,
2\pi\sigma_{\nu_p}.
\tag{B.12}
\]

Thus we expect that, for a given \(G_{\max}\), the penalties
in short fibers will increase with \(\sqrt{L}\) and will
be rather insensitive to $
|\lambda_1-\lambda_0|.
$ 
The \(Q\) penalty due to variations in pump power
has to be treated separately for the 1P- or the
2P-FOPA. We start with the 1P-FOPA.
\(\sigma_G\) is obtained from

\[
\sigma_G
\simeq
\left|
\frac{\partial G}{\partial P_1}
\right|
\sigma_{P_1},
\tag{B.13}
\]

and

\[
\frac{\partial G}{\partial P_1}
=
(G-1)\frac{x_0}{P_1}
+
\frac{x_0^2}{2}
\frac{\partial G}{\partial x}
\frac{\partial x}{\partial P_1}.
\tag{B.14}
\]

Thus we have

\[
\frac{\sigma_G}{G}
\simeq
\left[
2(G-1)
-
x_0^2 h(x)\,\Delta\beta\,L
\right]
\frac{\sigma_{P_1}}{P_1},
\tag{B.15}
\]

where $
h(x)=f(x)G(x)
$ is a monotonic decreasing
function of \(x\), with maximum value $
h(0)=\frac13.
$ The first term in Eq.~(B.15) varies little in the
region of interest $
(0<x<x_0)
$ and is essentially a
constant factor of 2. This term is important only
in the same spectral region where $
\Delta\beta\sim0
$ (which,
for $
\beta_{21}<0
$ and $
\beta_{41}<0,
$ this occurs only if $
\omega\simeq0),
$ i.e., around the central pump. If $
\Delta\beta=0
$ we have $
\frac{\sigma_G}{G}
=
2\frac{\sigma_{P_1}}{P_1}
(1+x_0^2)
\simeq
2\frac{\sigma_{P_1}}{P_1},
$ which can be used to estimate the minimum
\(Q\) penalty.

The maximum penalty occurs in the wings of the gain
spectrum, where the second term in Eq.~(B.15)
dominates. Within the region of interest, this term
has the largest value at $
x=0,
$ where $
\Delta\beta=-4\gamma P_1,
$ and \[
\frac{\sigma_G}{G}
\simeq
\left|
1+
\frac{2x_0^2}{x_0^2+4}
\frac{4\gamma P_1}{P_1}
\right|.
\tag{B.16}
\]

Thus, the analytical modeling of the 1P-FOPA
indicates that $
\Delta Q,
$ due to either pump RIN or
pump phase modulation, is minimized for signal
wavelengths where $
\Delta\beta\simeq0,
$ near the pump,
and is maximized for $
\Delta\beta\sim-4\gamma P_1
$ (i.e., near –
but not exactly – the inflection points at the outer
slopes of the gain spectrum).

For the 2P-FOPA case, the
\(\sigma_G\) due to uncorrelated stochastic variations of
\(P_1\) and
\(P_2\) is given by \[
\sigma_G
=
\sqrt{
\sigma_{G_1}^2+\sigma_{G_2}^2
},
\tag{B.17a}
\]

\[
\frac{\sigma_G}{G}
=
\left[
\left(
\frac{G-1}{G}
\right)
\left(
1+
\frac{3x_0^2}{L}\,
x_0\Delta\beta
\right)
\left(
1-
\frac{\sinh(2x)}{4g\,\sinh^2(x)}
\frac{L}{4x_0^2}
\right)
\right]
\frac{\sigma_{P_1}}{P_1}.
\tag{B.17b}
\]

---

\section*{ Appendix C. Pump phase modulation with a pseudo-random bit sequence}

Binary phase shift keyed phase modulation is
commonly employed in actual FOPAs to suppress
the stimulated Brillouin scattering
[18,21,31].
Fig.~C.1(a) shows a typical broadened pump spectrum
(measured with 0.01 nm resolution bandwidth)
that was obtained using our LiNbO\(_3\) Mach–Zehnder
phase modulator driven by a 
\(10~\mathrm{Gb/s}\),
\(2^{31}-1\) 
pseudorandom bit sequence.

In general, the spectral characteristic of the
broadened pump depends on the particular phase
modulator (PM) and the RF amplifier used to
drive the PM; however, the spectrum shown in
Fig.~C.1(a) is a representative one of a pump
phase modulated with a PRBS (see [21]).

To investigate the effect of the PRBS phase
modulation, we solved Eq.~(8) for two sets of
400 pump frequencies that were chosen in order
to emulate the pump broadening of Fig.~C.1(a).

In Figs.~C.1(b) and (c) we show the histograms of
these two sets of 400 frequency values. Even though the spectral distributions are different,
we choose the two sets of pump frequencies in
order to have identical standard deviation: $
\sigma_{\nu}=2.45~\mathrm{GHz}.
$ Fig.~C.2 shows the
\(Q\)-penalty spectra resulting from these two sets
of frequencies. Dashed and continuous lines correspond to
Fig.~C.1(b) and (c), respectively. Note that, in spite of the two spectral distributions
having identical $
\sigma_{\nu},
$ the \(Q\) penalties are larger at the first peak for
the broader distribution. For comparison we also plot in dotted lines the
\(Q\)-penalty obtained using a uniform distribution
with $
\sigma_{\nu}=2.45~\mathrm{GHz}.
$ The conclusion from Fig.~C.2 is that, for a given
standard deviation, the \(Q\) penalties increase as
long as the spectrum is broader.

%%%%%%%%%%%%%%%%%%%%%%%%%%%%%%%%%%%%%%%%%%%%%%%%%%%%%%%%%%%%%%

\section*{Appendix D. Random variations of \(\lambda_0\)}

Eq.~(8) was numerically solved by dividing the
fiber in 3000 segments of equal length, each one
having a different value of the zero dispersion
wavelength. The
\(\lambda_0\)
in each step was modeled as being $
\lambda_0(z)
=
1567~\mathrm{nm}
+
\delta_{\lambda_0}(z)
+
\Delta_{\lambda_0}(z),
$ where $
\delta_{\lambda_0}(z)
$ and $
\Delta_{\lambda_0}(z)
$ are the short- and long-length-scale
fluctuations, respectively.

To construct the short-length-scale fluctuation, $
\delta_{\lambda_0}(z), $ we generated 3000 numbers $
n_k,
\qquad
(k=1,2,\ldots,3000),
$ from a random Gaussian process with zero mean
and standard deviation $
1~\mathrm{nm}.
$ We defined a set of values $
\delta_{\lambda_0}(z_k)
=
\left[
n_k
+
0.7\,n_{k-1}
+
0.6\,n_{k-2}
+
0.5\,n_{k-3}
+
0.4\,n_{k-4}
\right]\,s,
$ for $
k\ge5,
$ and $
\delta_{\lambda_0}(z_k)
=
s\,n_k,
$ for $
k<5,
$ where $
s
$ is chosen to obtain the desired standard
deviation. The resulting distribution has zero mean and
autocorrelation length $
l_c \cong 11~\mathrm{m}.
$ In most cases, the standard deviation was $
0.02~\mathrm{nm}
$ (i.e., $
s\sim\frac{1}{30}
$ ). This value for the standard deviation of the
short-length-scale fluctuation is realistic in
typical commercial dispersion-shifted fibers but
not in highly nonlinear fibers, for which the
fluctuation is larger by two orders of magnitude. In some cases we wanted a standard deviation
smaller than $
0.02~\mathrm{nm}.
$ To model the long-length-scale $
\lambda_0
$ variations we followed a similar approach: 10 numbers, $
p(j),
$ were generated from a random Gaussian process with zero mean and standard deviation
$
1~\mathrm{nm}.
$ The correlations were introduced by defining $
\Delta\lambda_0(j)
=
p(j)
+
0.8\,p(j-1),
\qquad j>2,
$
and
$
\Delta\lambda_0(1)=p(1).
$ By varying the scaling factor we can vary the
standard deviation of $\Delta\lambda_0(j).$ The resulting slow fluctuation has a constant value
for every $0.68~\mathrm{km}$ segment of fiber (i.e., for 300 values of
\(\lambda_0\)), zero mean, and correlation length $1.1~\mathrm{km}.$

In Fig.~D.1 we show a typical map of the resulting
\(\lambda_0(z)\) variation, where both the long- and
the short-length-scale variations of
\(\lambda_0\)
can be observed.

Using this procedure we simulated random
variations of \(\lambda_0\) in both 1P-FOPAs and 2P-FOPAs.
For the simulation of Fig.~3(b) we considered only
the short-length-scale fluctuations.

%%%%%%%%%%%%%%%%%%%%%%%%%%%%%%%%%%%%%%%%%%%%%%%%%%%%%%%%%%%%%%

\section*{References}

[1] J. Hansryd, P.A. Andrekson, IEEE Photon. Technol. Lett.
11 (2001) 194.

[2] K. Inoue, J. Lightwave Technol. 20 (2002) 969.

[3] K.K.Y. Wong, M.E. Marhic, K. Uesaka,
L.G. Kazovsky,
IEEE Photon. Technol. Lett. 14 (2002) 911.

[4] R. Tang, P.L. Voss, J. Lasri,
P. Devgan, P. Kumar,
Opt. Lett. 29 (2004).

[5] M.C. Ho, M.E. Marhic,
K.K.Y. Wong,
L.G. Kazovsky,
J. Lightwave Technol. 20 (2002) 469.

[6] J.L. Blows, S.E. French,
Opt. Lett. 27 (2002) 491.

[7] L. Provino,
A. Mussot,
E. Lantz,
C. Simonneau,
T. Sylvestre,
H. Maillotte,
J. Opt. Soc. Am. B 20 (2003) 1532.

[8] J.M. Chavez Boggio,
S. Tenenbaum,
H.L. Fragnito,
J. Opt. Soc. Am. B 18 (2001) 1428.

[9] M. Farahmand,
M. de Sterke,
Opt. Express 12 (2004) 136.

[10] J.D. Marconi,
J.M. Chavez Boggio,
H.L. Fragnito,
Electron. Lett. 40 (2004) 1213.

[11] F. Yaman,
Q. Lin,
G.P. Agrawal,
IEEE Photon. Technol. Lett. 16 (2004) 431.

[12] G. Kalogerakis,
M.E. Marhic,
K.K.Y. Wong,
L.G. Kazovsky,
CLEO (2004), Paper CFAS.

[13] K.K.Y. Wong,
K. Shimizu,
M.E. Marhic,
K. Uesaka,
G. Kalogerakis,
L.G. Kazovsky,
Opt. Lett. 28 (2003) 692.

[14] W. Zhang,
C. Wang,
J. Shu,
C. Jiang,
W. Hu,
IEEE Photon. Technol. Lett. 16 (2004) 1652.

[15] J.M. Chavez Boggio,
P. Dainese,
F. Karlsson,
H.L. Fragnito,
IEEE Photon. Technol. Lett. 15 (2003) 1528.

[16] C.J. McKinstrie,
S. Radic,
A. Chraplyvy,
IEEE J. Sel. Top. Quantum Electron. 8 (2002) 538.

[17] K. Uesaka,
K.K.Y. Wong,
M.E. Marhic,
L.G. Kazovsky,
IEEE J. Sel. Top. Quantum Electron. 8 (2002) 560.

[18] S. Radic,
C.J. McKinstrie,
R.M. Jopson,
J.C. Centanni,
A.R. Chraplyvy,
C.G. Jørgensen,
K. Brar,
C. Headley,
IEEE Photon. Technol. Lett. 15 (2003) 673.

[19] K.K.Y. Wong,
K. Shimizu,
K. Uesaka,
G. Kalogerakis,
M.E. Marhic,
L.G. Kazovsky,
IEEE Photon. Technol. Lett. 15 (2003) 1707.

[20] M.E. Marhic,
K.K.Y. Wong,
G. Kalogerakis,
L.G. Kazovsky,
Opt. Photon. News (2004) 22.

[21] S. Radic,
C.J. McKinstrie,
R.M. Jopson,
A.H. Gnauck,
J.C. Centanni,
A.R. Chraplyvy,
IEEE Photon. Technol. Lett. 16 (2004) 548.

[22] T. Sakamoto,
A. Okada,
O. Moriwaki,
M. Matsuoka,
K. Kikuchi,
J. Lightwave Technol. 22 (2004) 874.

[23] A. Mussot,
A. Durécu-Legrand,
E. Lantz,
C. Simonneau,
D. Bayart,
H. Maillotte,
T. Sylvestre,
IEEE Photon. Technol. Lett. 16 (2004) 1289.

[24] A. Legrand,
S. Lanne,
C. Simonneau,
D. Bayart,
A. Mussot,
E. Lantz,
T. Sylvestre,
H. Maillotte,
Optical Fiber Communication Conference (OFC 2004),
paper TuK2, OSA Technical Digest,
Los Angeles, California, USA,
February 2004,
pp. 22--27.

[25] P.O. Hedekvist,
P.A. Andrekson,
J. Lightwave Technol. 17 (1999) 74.

[26] M.N. Islam,
O. Boyraz,
IEEE J. Sel. Top. Quantum Electron. 8 (2002) 527.

[27] P. Kylemark,
P.O. Hedekvist,
M. Karlsson,
P.A. Andrekson,
J. Lightwave Technol. 22 (2004) 409.

[28] C.R.S. Fludger,
V. Handerek,
R.J. Mears,
J. Lightwave Technol. 19 (2001) 1140.

[29] M. Yu,
C.J. McKinstrie,
G.P. Agrawal,
J. Opt. Soc. Am. B 12 (1995) 1126.

[30] S.K. Korotky,
P.B. Hansen,
L. Eskildsen,
J.J. Veselka,
Tech. Dig. IOOC'95,
vol. 2,
Paper WD2,
p. 110.

[31] T. Tanemura,
H.C. Lim,
K. Kikuchi,
IEEE Photon. Technol. Lett. 13 (2001) 1328.

[32] R.M. Jopson,
U.S. Patent 5,386,314 (1994).

[33] K.K.Y. Wong,
M.E. Marhic,
L.G. Kazovsky,
IEEE Photon. Technol. Lett. 15 (2003) 33.

[34] J.M. Chavez Boggio,
S. Tenenbaum,
H.L. Fragnito,
Optical Fiber Communication Conference (OFC 2001),
OSA Technical Digest,
paper WDD24.

[35] F. Yaman,
Q. Lin,
S. Radic,
G.P. Agrawal,
IEEE Photon. Technol. Lett. 16 (2004) 1292.

[36] M. Karlsson,
J. Opt. Soc. Am. B 15 (1998) 2269.

[37] T. Okuno,
M. Tanaka,
M. Hirano,
T. Kato,
M. Shigematsu,
M. Onishi,
Proc. European Conference on Optical Communication (ECOC),
Rimini, Italy,
paper We4.P.29,
September 2003,
p. 614.

[38] N. Kuwaki,
M. Ohashi,
J. Lightwave Technol. 8 (1990) 1476.

[39] A. Mussot,
E. Lantz,
T. Sylvestre,
H. Maillotte,
A. Durécu,
C. Simonneau,
D. Bayart,
European Conference on Optical Communication (ECOC 2004),
paper Tu3.3.7,
Sweden,
September 2004.

[40] A. Legrand,
C. Simonneau,
D. Bayart,
A. Mussot,
E. Lantz,
T. Sylvestre,
H. Maillotte,
Optical Amplifiers and their Applications (OAA 2003),
Otaru, Japan,
July 2003.

\section*{Captions}

\begin{figure}[H]
\centering
\includegraphics[width=\textwidth]{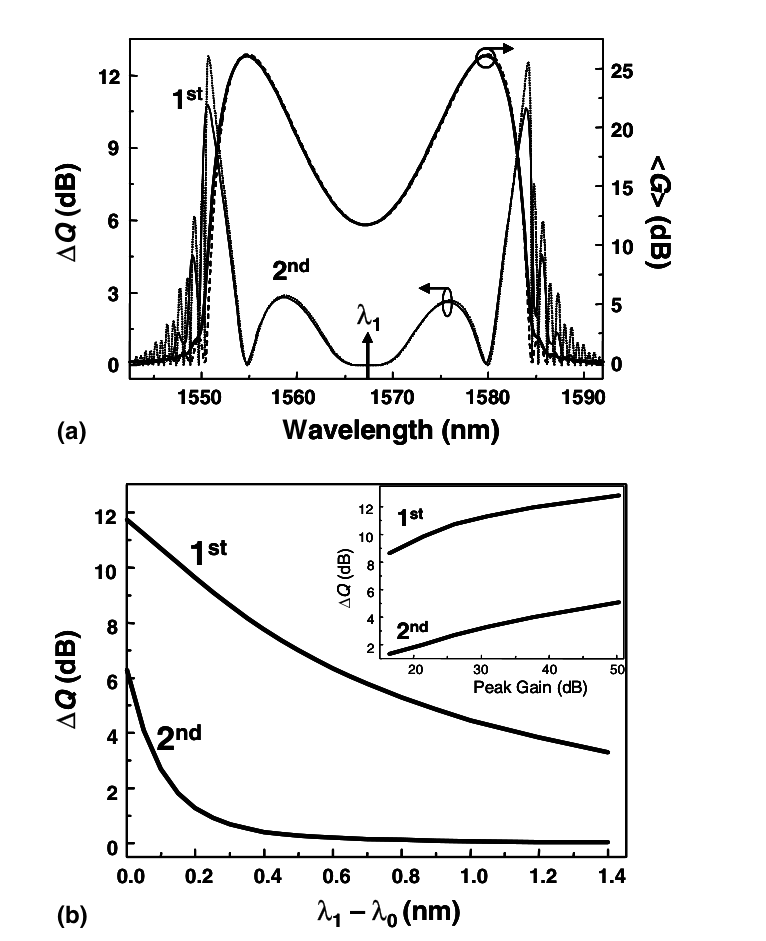}
\label{fig:1}
\end{figure}

\noindent
\textbf{Figure 1.}

(a) $\Delta Q$ and $\langle G\rangle$ as functions of
$\lambda_s$ for $
\lambda_1-\lambda_0=0.1~\mathrm{nm}.
$ Solid lines correspond to numerical calculations.
Dotted lines represent the analytical prediction of
$\Delta Q$, while segmented lines represent the analytical gain.
The largest penalties occur at the outer slopes of the gain spectrum.

(b) $\Delta Q$ as a function of
$\lambda_1-\lambda_0$.
As the pump wavelength approaches the zero-dispersion wavelength,
the $Q$ penalties increase.
The inset shows $\Delta Q$ as a function of the peak parametric gain.
The baseline quality factor used in all calculations is $Q_i=11.$

\bigskip

\begin{figure}[H]
\centering
\includegraphics[width=\textwidth]{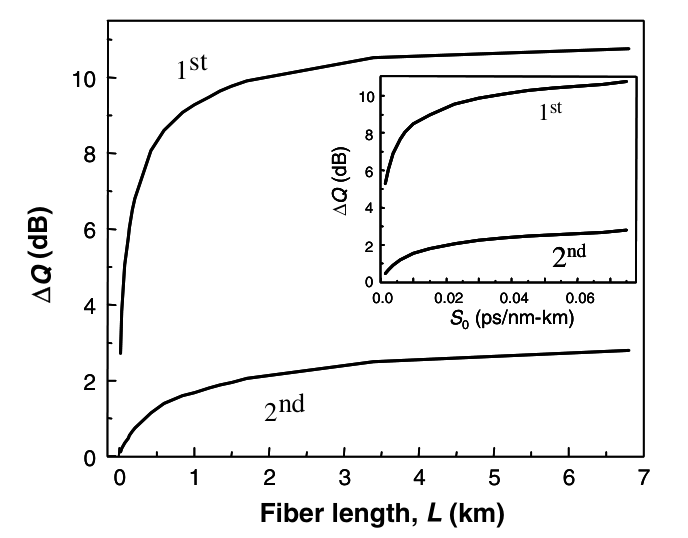}
\label{fig:2}
\end{figure}

\noindent
\textbf{Figure 2.}

$\Delta Q$ at the first and second peaks as a function of the fiber
length $L$. Inset: $\Delta Q$ at the first and second peaks as a function of the
fiber dispersion slope $S_0$.

\bigskip

\begin{figure}[H]
\centering
\includegraphics[width=\textwidth]{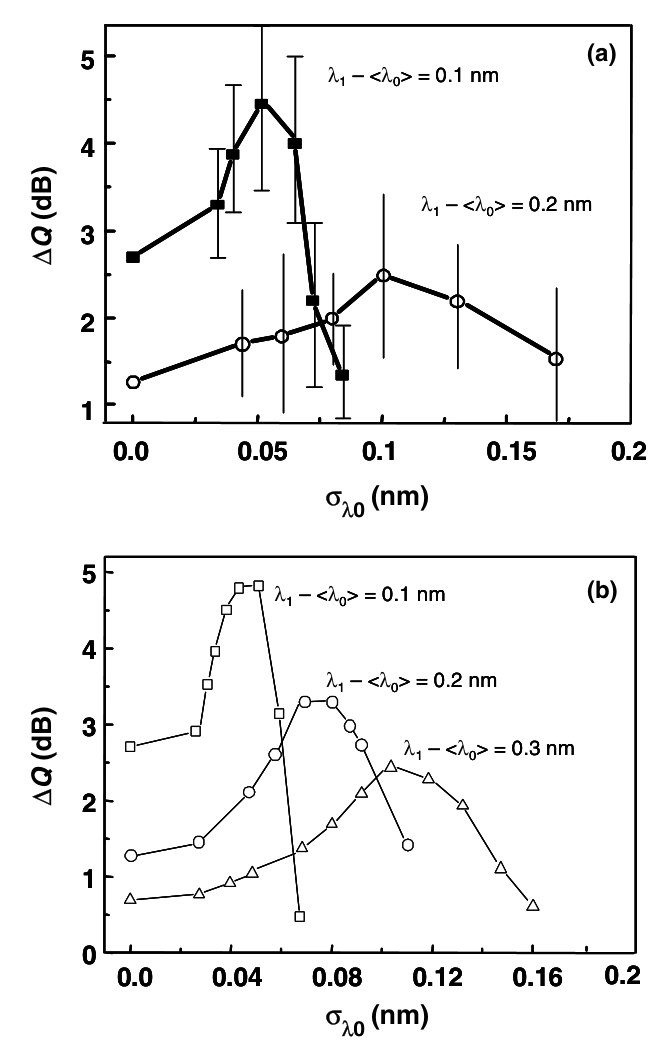}
\label{fig:3}
\end{figure}

\noindent
\textbf{Figure 3.}

(a) Second-peak $Q$ penalty as a function of $\sigma_{\lambda_0}$ for fibers with random longitudinal
variations of $\lambda_0(z)$ as described in Appendix~D. The error bars represent the spread obtained from the 15 simulated
fiber realizations.

(b) Second-peak $Q$ penalty as a function of $\sigma_{\lambda_0}$ for a fiber composed of three sections having
increasing values of $\lambda_0$. For $\lambda_1-\langle\lambda_0\rangle>0.3~\mathrm{nm}$,
the simulations produced essentially the same dependence of $\Delta Q$ on $\sigma_{\lambda_0}$.

\bigskip

\begin{figure}[H]
\centering
\includegraphics[width=\textwidth]{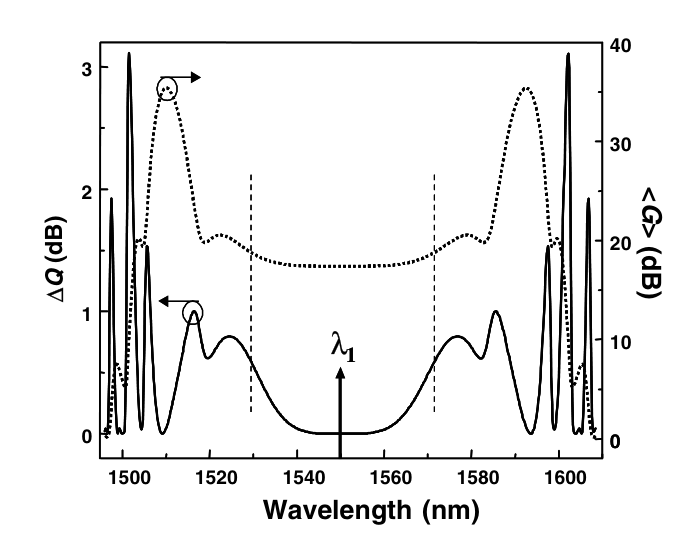}
\label{fig:4}
\end{figure}

\noindent
\textbf{Figure 4.}

Average gain $\langle G\rangle$ (dotted line) and
$Q$ penalty (solid line) for a dispersion-tailored fiber whose
parameters are listed in Table~1.
The segmented vertical lines indicate the wavelength interval where
the gain ripple is less than approximately $1.6~\mathrm{dB}$ and the corresponding
$Q$ penalty remains below $0.6~\mathrm{dB}$.

\bigskip

\begin{figure}[H]
\centering
\includegraphics[width=\textwidth]{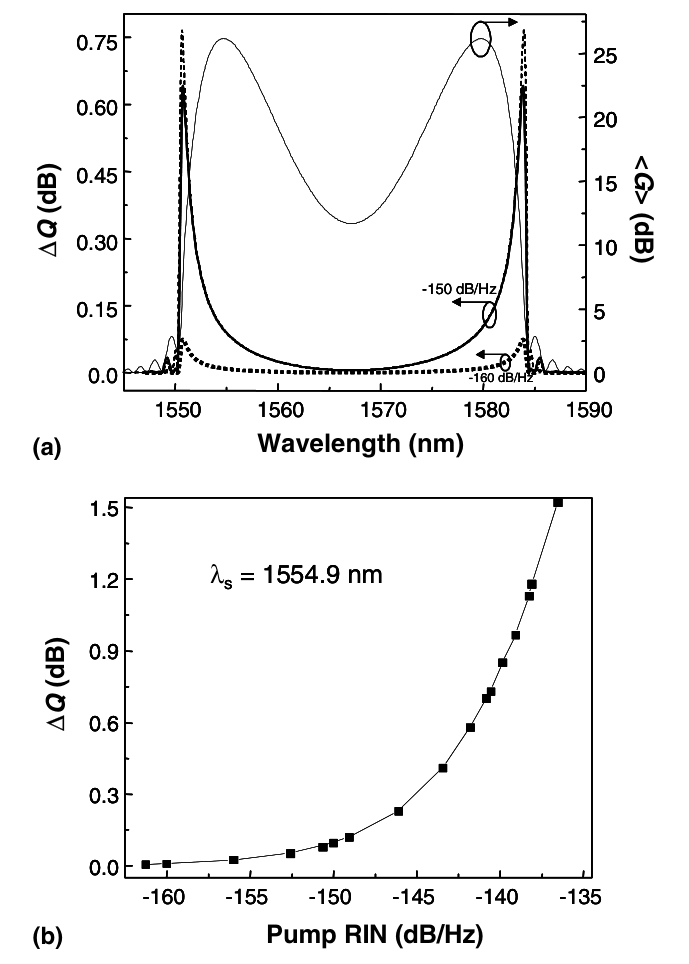}
\label{fig:5}
\end{figure}

\textbf{Fig. 5.}
(a) $\Delta Q$ and $\langle G\rangle$ as a function of $\lambda_s$
for $-160$ dB/Hz (dotted lines) and $-150$ dB/Hz (solid lines)
pump RIN.
(b) $\Delta Q$ value for the signal wavelength at the gain peak
as a function of pump RIN.
The FOPA parameters are identical to those in Fig.~2(a), and no
phase modulation is applied to the pump.

\bigskip

\begin{figure}[H]
\centering
\includegraphics[width=\textwidth]{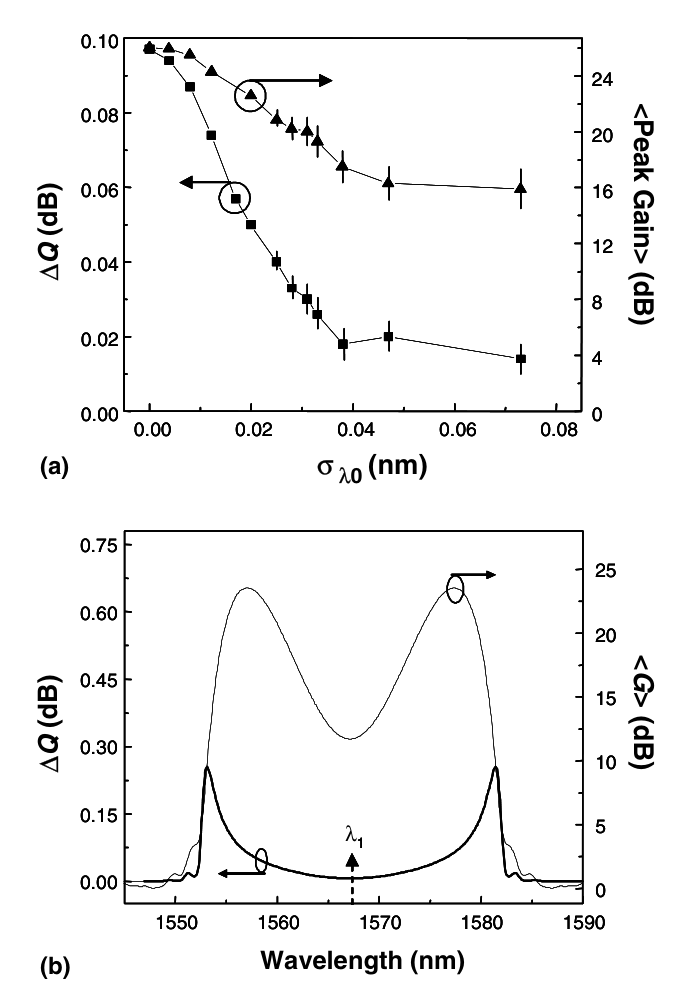}
\label{fig:6}
\end{figure}

\textbf{Fig. 6.}
(a) $\Delta Q$ calculated at the gain peak as a function of
$\sigma_{\lambda_0}$.
For comparison, the evolution of the peak gain is shown by triangles.
The bars indicate the range of variation of $\Delta Q$ and
$\langle G_{\mathrm{peak}}\rangle$.
(b) $\Delta Q$ and $\langle G\rangle$ as a function of $\lambda_s$
for $\sigma_{\lambda_0}=0.021~\mathrm{nm}$
(the other 1P-FOPA parameters are identical to those of Fig.~4(a)).

\bigskip

\begin{figure}[H]
\centering
\includegraphics[width=\textwidth]{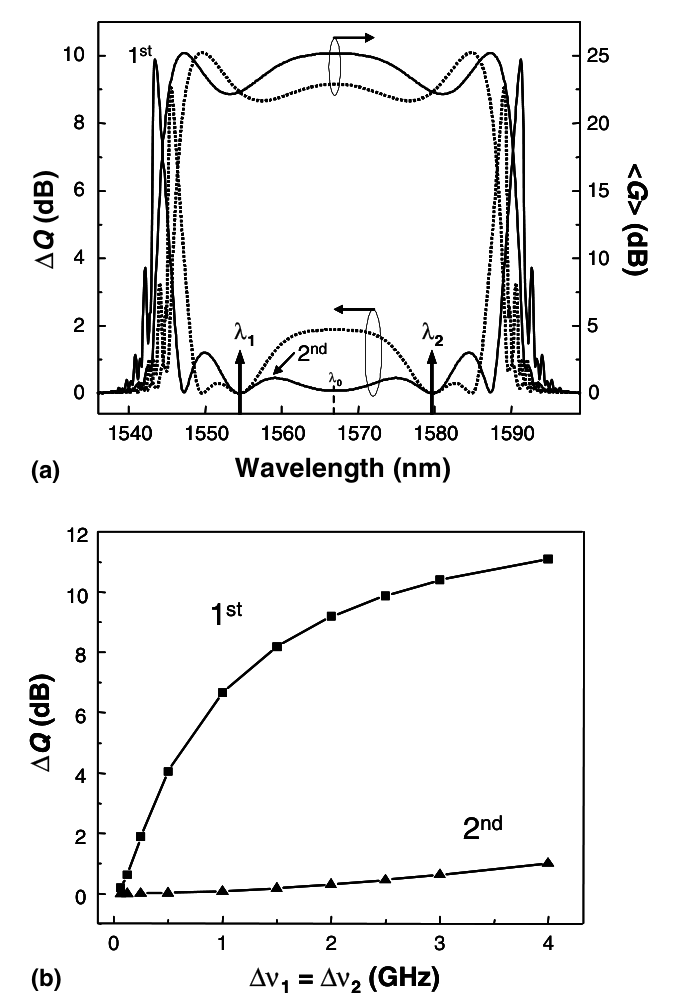}
\label{fig:7}
\end{figure}

\textbf{Fig. 7.}
(a) $\Delta Q$ and $\langle G\rangle$ as a function of $\lambda_s$
for two positions of $\lambda_1$ and $\lambda_2$.
(b) Evolution of $\Delta Q$, calculated at the first and second peaks,
as a function of the pump broadening
$\Delta\nu_1=\Delta\nu_2$.

\bigskip

\begin{figure}[H]
\centering
\includegraphics[width=\textwidth]{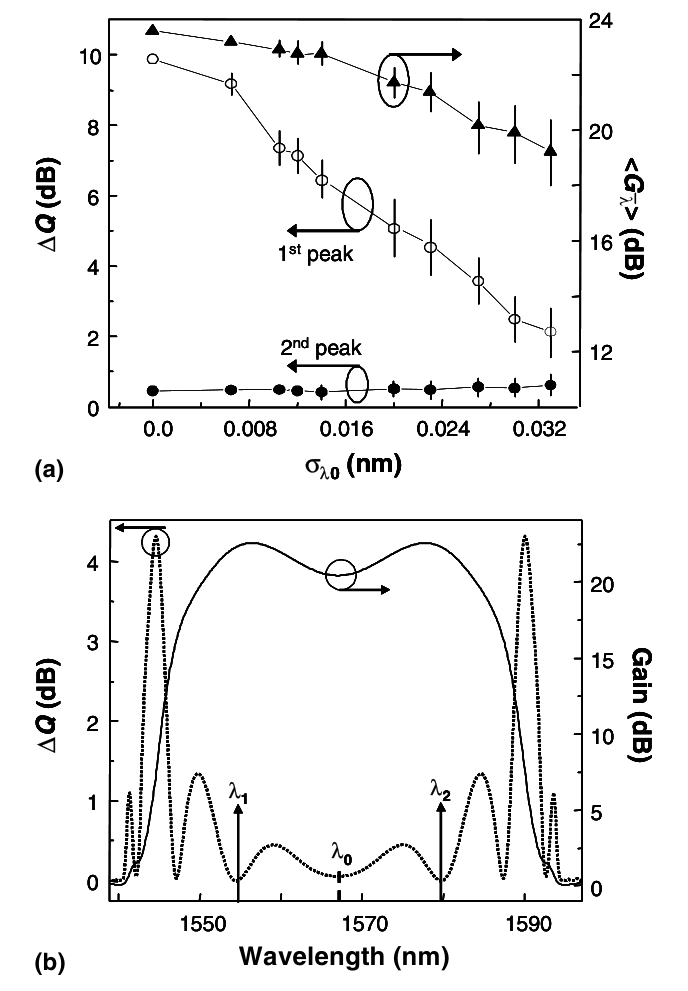}
\label{fig:8}
\end{figure}

\textbf{Figure 8.}

(a) First- and second-peak $Q$ penalties as functions of
$\sigma_{\lambda_0}$.
The triangles represent the average gain evaluated at the center of the
gain spectrum.
Error bars correspond to the statistical spread obtained from the
ensemble of simulated fibers.

(b) Average gain spectrum and corresponding
$Q$ penalty for $
\sigma_{\lambda_0}
=
0.021~\mathrm{nm},
$ using the same longitudinal distribution of
$\lambda_0(z)$
employed in Fig.~6(b).

\bigskip

\begin{figure}[H]
\centering
\includegraphics[width=\textwidth]{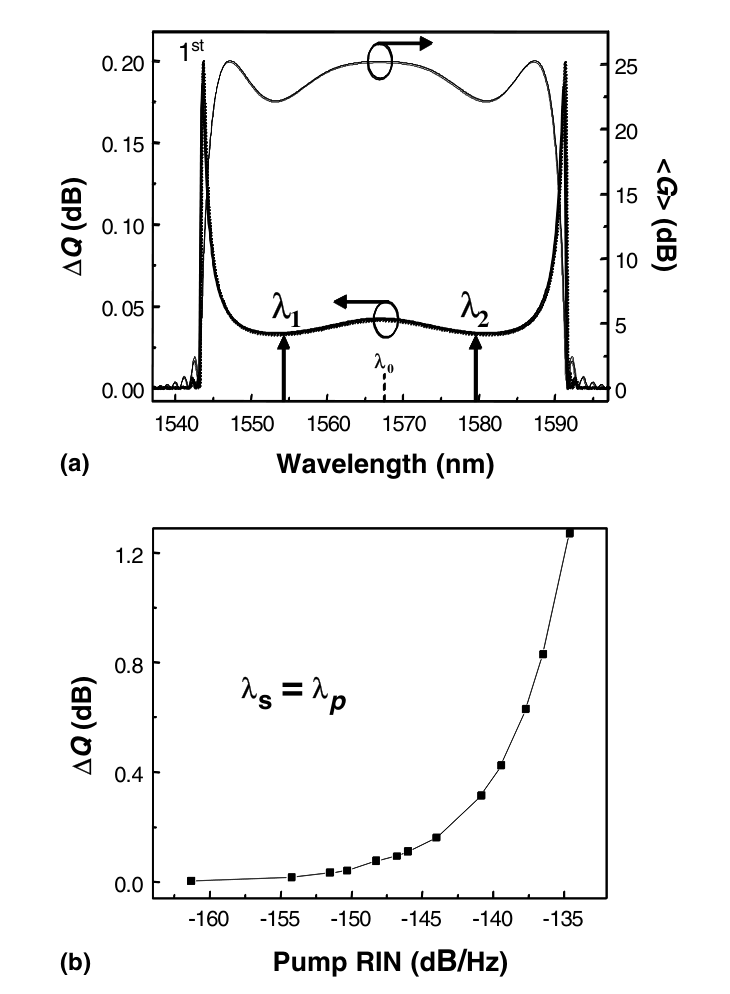}
\label{fig:9}
\end{figure}

\textbf{Fig. 9.}
(a) $\Delta Q$ and $\langle G\rangle$ as a function of $\lambda_s$
for $\mathrm{RIN}=-150~\mathrm{dB/Hz}$.
Solid line: numerical calculation.
Dotted line: analytical calculation.
(b) Evolution of $\Delta Q$, calculated at $\lambda_s=\lambda_p$,
as a function of the pump RIN.

\bigskip

\begin{figure}[H]
\centering
\includegraphics[width=\textwidth]{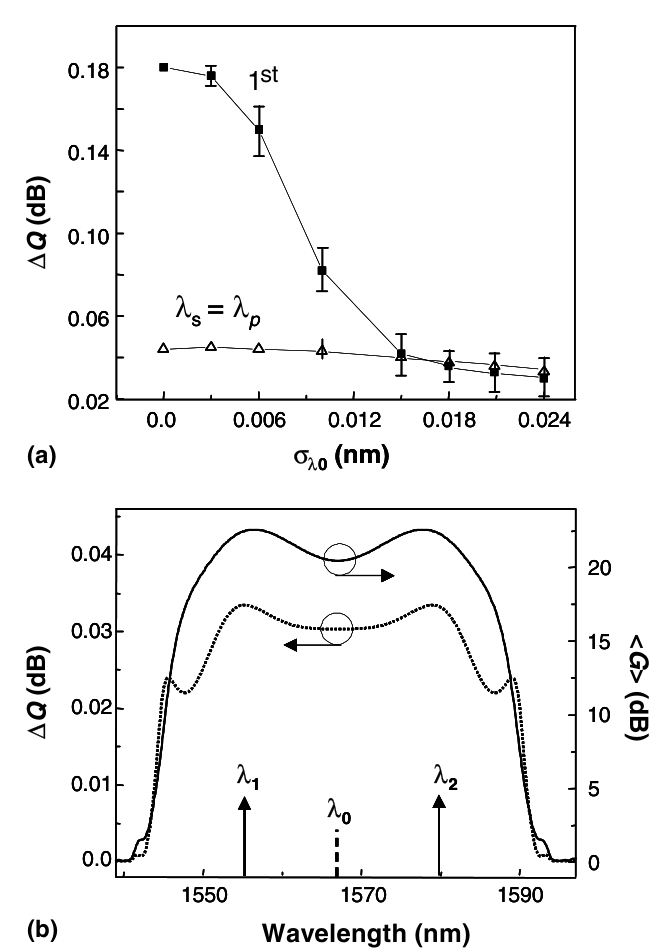}
\label{fig:10}
\end{figure}

\textbf{Figure 10.}

(a) First-peak $\Delta Q$ (solid squares) and $\Delta Q$ evaluated at
$\lambda_s=\lambda_p$ (open triangles) as functions of
$\sigma_{\lambda_0}$. The error bars indicate the statistical spread obtained from the
numerical simulations.

(b) Spectra of
$\Delta Q$
and
$\langle G\rangle$
for $
\sigma_{\lambda_0}=0.021~\mathrm{nm},
$ using the same longitudinal variation of
$\lambda_0(z)$
as in Fig.~6(b).

\bigskip

\begin{figure}[H]
\centering
\includegraphics[width=\textwidth]{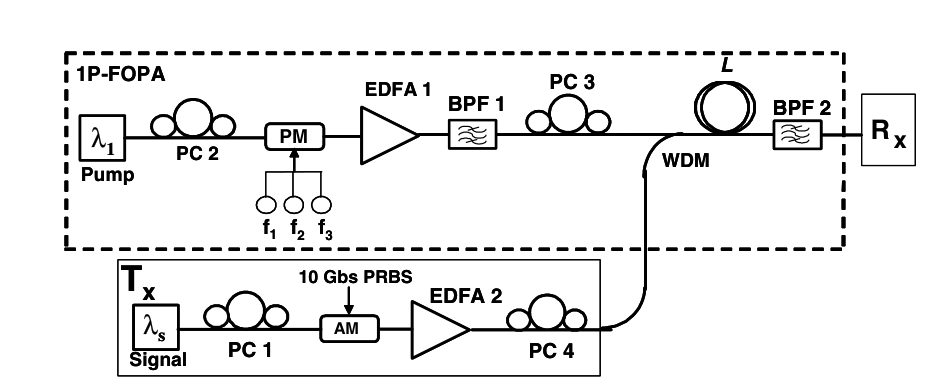}
\label{fig:11}
\end{figure}

\textbf{Figure 11.}

Experimental setup used for the single-pump FOPA experiments. For the on/off gain measurements, the optical
signal-to-noise ratio at the receiver input was approximately $
33~\mathrm{dB}.
$ The baseline quality factor, $
Q_i,
$ measured for different signal wavelengths, ranged between $
11
\quad\text{and}\quad
12.
$

\bigskip

\begin{figure}[H]
\centering
\includegraphics[width=\textwidth]{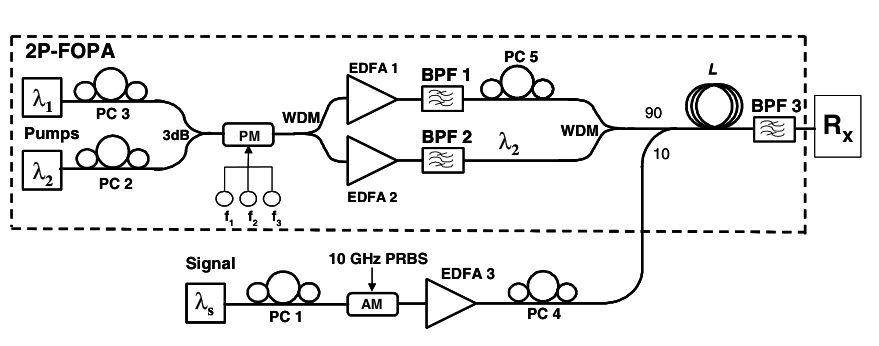}
\label{fig:12}
\end{figure}

\textbf{Figure 12.}

Eye diagrams of the received signal as a function of the pump
detuning. (a) Back-to-back signal, $\lambda_1-\langle\lambda_0\rangle=0.25~\mathrm{nm}$.
(b) $\lambda_1-\langle\lambda_0\rangle=0.25~\mathrm{nm}$.
(c) $\lambda_1-\langle\lambda_0\rangle=0.20~\mathrm{nm}$. (d) $\lambda_1-\langle\lambda_0\rangle=0.16~\mathrm{nm}$.

\bigskip

\begin{figure}[H]
\centering
\includegraphics[width=\textwidth]{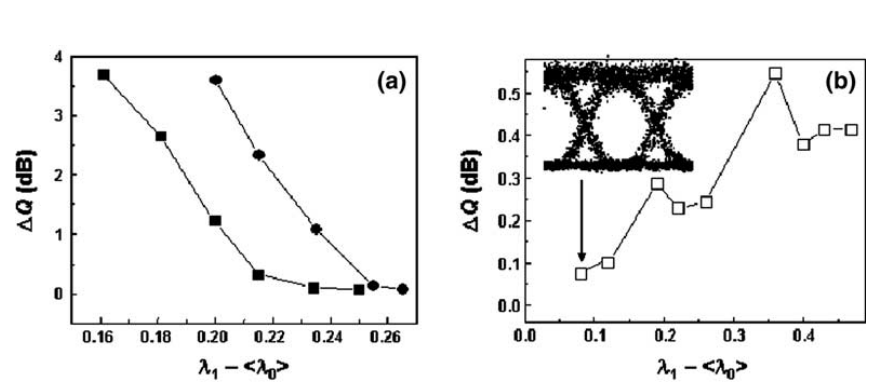}
\label{fig:13}
\end{figure}

\textbf{Figure 13.}

Measured and calculated $\Delta Q$ as a function of the signal wavelength.
The agreement between theory and experiment confirms the validity of
the numerical model developed in this work.

\bigskip

\begin{figure}[H]
\centering
\includegraphics[width=\textwidth]{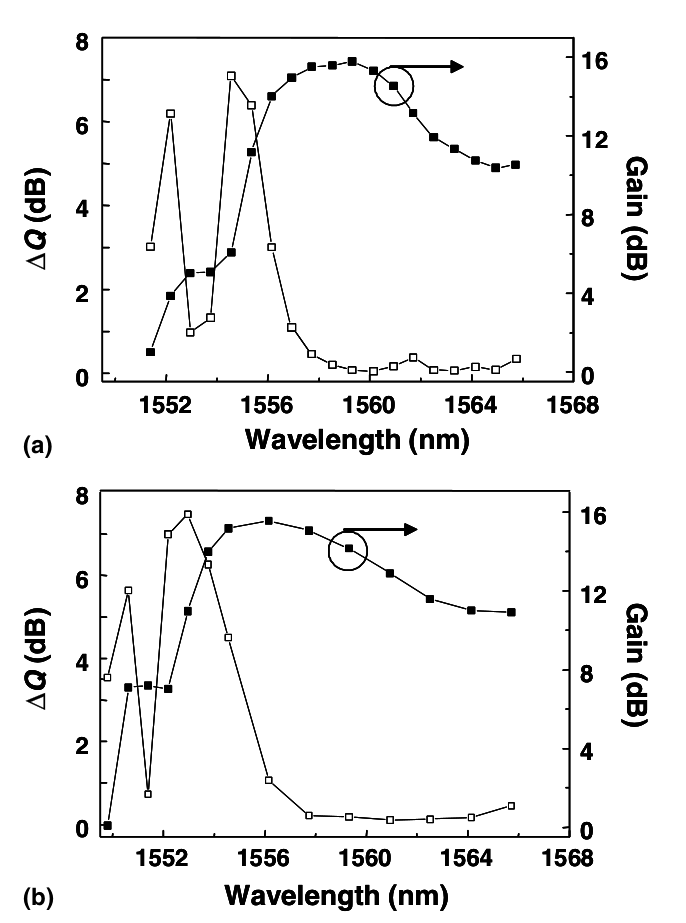}
\label{fig:14}
\end{figure}

\textbf{Fig. 14.}
$\Delta Q$ (hollow squares) and $G$ (solid squares) as a function
of $\lambda_s$ for $P_1\simeq330~\mathrm{mW}$ with fiber A.
(a) Band-pass filter after the booster EDFA.
(b) Band-pass filter before the booster EDFA.

\bigskip

\begin{figure}[H]
\centering
\includegraphics[width=\textwidth]{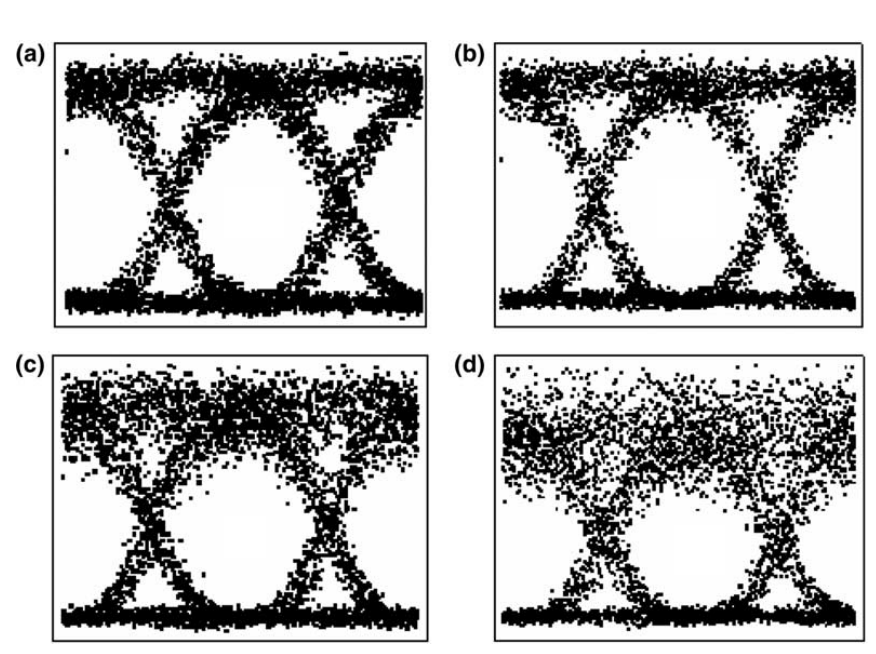}
\label{fig:15}
\end{figure}

\textbf{Figure 15.}

Experimental setup for 2P-FOPA experiments.

\bigskip

\begin{figure}[H]
\centering
\includegraphics[width=\textwidth]{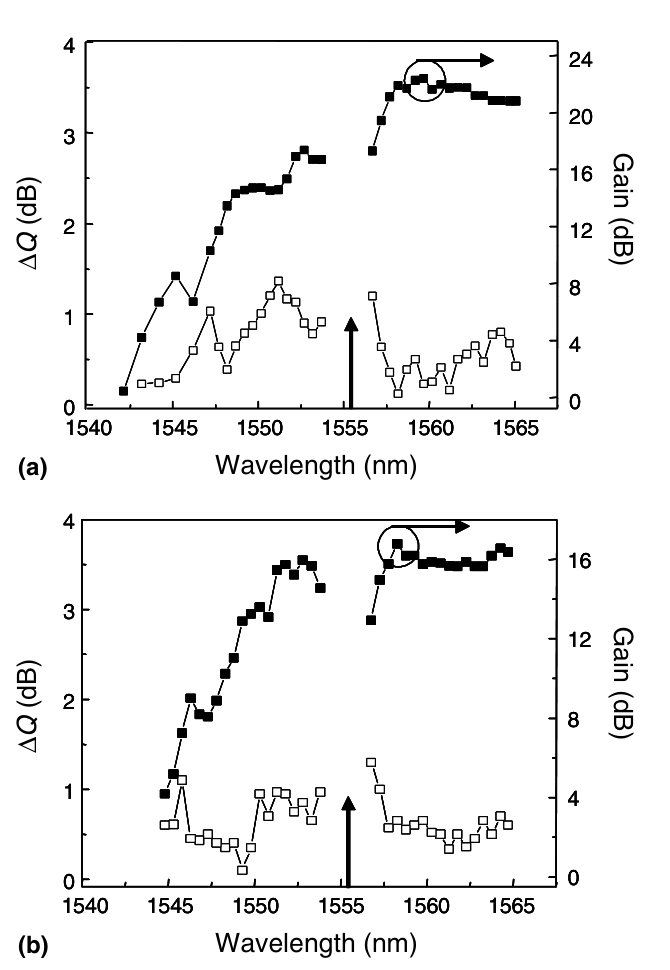}
\label{fig:16}
\end{figure}

\textbf{Figure 16.}

$\Delta Q$
(hollow squares) and
$\langle G\rangle$
(solid squares) as a function of $
\lambda_s.
$
(a) $
P_1 \approx P_2 = 24~\mathrm{dBm},
$
(b) $
P_1 \approx P_2 = 23~\mathrm{dBm}.
$

\bigskip

\begin{figure}[H]
\centering
\includegraphics[width=\textwidth]{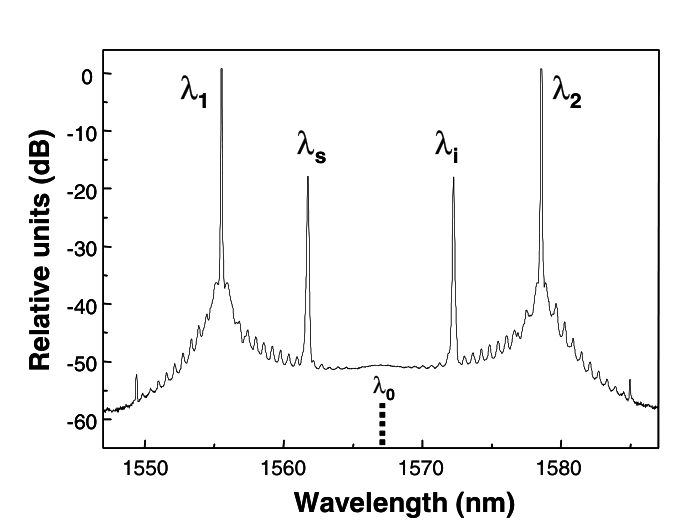}
\label{fig:17}
\end{figure}

\textbf{Figure 17.}

Typical spectrum at the output of the FOPA
(with $
G \approx 16~\mathrm{dB}
$ from Fig.~16(b)). Note that the OSNR is approximately $
32~\mathrm{dB}.$ 

\bigskip
\begin{figure}[H]
\centering
\includegraphics[width=\textwidth]{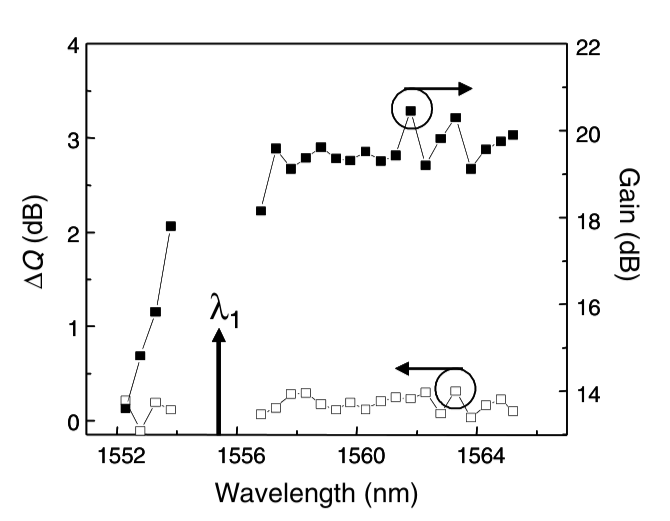}
\label{fig:18}
\end{figure}

\textbf{Figure 18.}

$\Delta Q$ (hollow squares) and
$\langle G\rangle$ (solid squares) as a function of
$\lambda_s$.
The conversion efficiency is approximately $
20\%,
$ which results in the best FOPA performance.

\bigskip

\begin{figure}[H]
\centering
\includegraphics[width=\textwidth]{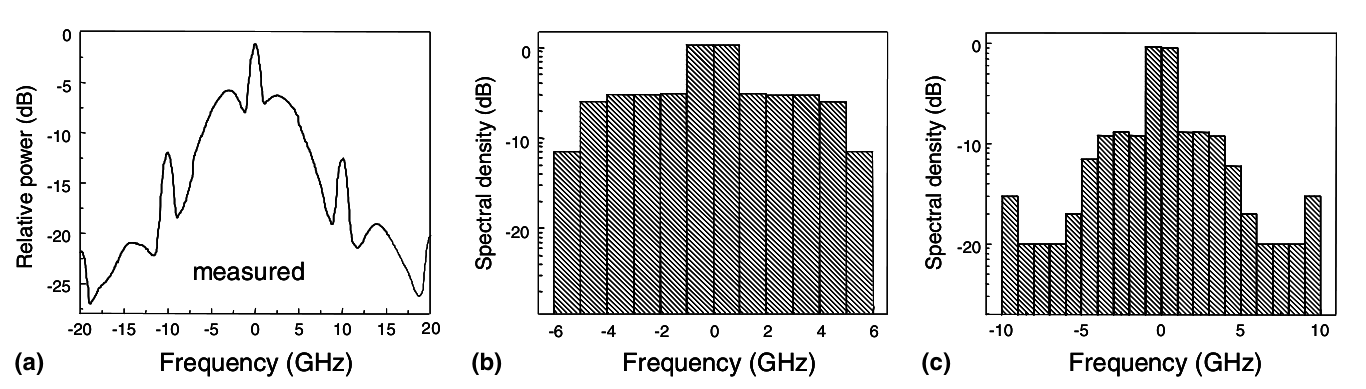}
\label{fig:C1}
\end{figure}

\textbf{Figure C.1.}

(a) Measured pump spectrum using a
\(0.01~\mathrm{nm}\)
resolution optical spectrum analyzer.

(b) and (c) Histograms of the 400 pump frequency
values.

Note that distribution (c) is extended over a
broader range than distribution (b).

However, both distributions have identical standard
deviation.

\bigskip

\begin{figure}[H]
\centering
\includegraphics[width=\textwidth]{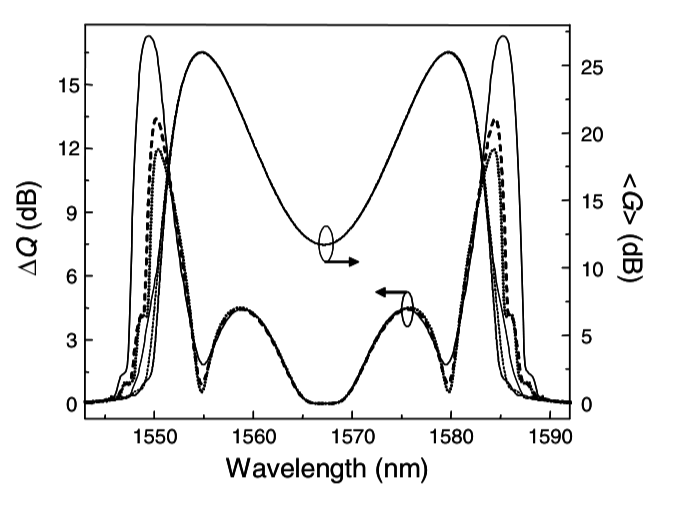}
\label{fig:C2}
\end{figure}

\textbf{Figure C.2.}

\(Q\) penalties using pump frequency distributions
from Fig.~C.1(c) (dashed line) and (c)
(continuous line).

For comparison (dotted line) we graphed the
\(Q\) penalty using a uniform distribution with
identical standard deviation.

\bigskip

\begin{figure}[H]
\centering
\includegraphics[width=\textwidth]{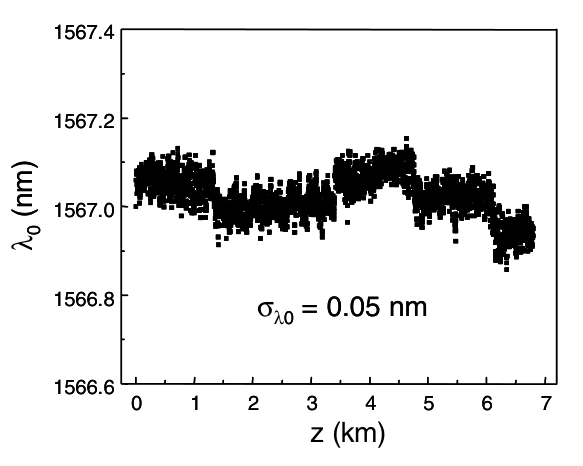}
\label{fig:D1}
\end{figure}

\textbf{Fig. D.1.}
Typical map of the longitudinal variation of
$\lambda_0$.
Both short- and long-length-scale fluctuations of
$\lambda_0$ can be observed.

\end{document}